\documentclass[aps,nofootinbib]{revtex4}

\usepackage{amsmath,amssymb,graphicx,float}
\usepackage{color}
\usepackage[normalem]{ulem}

\newcommand{\mathsym}[1]{{}}
\newcommand{\unicode}[1]{{}}

\begin{document}

\title{A numerical recipe for the computation of stationary stochastic processes' autocorrelation function}
\author{S. Miccich\`e}
\affiliation{Dipartimento di Fisica e Chimica - Emilio Segr\`e, Universit\`a degli Studi di Palermo, Viale delle Scienze, Ed. 18, 90128, Palermo, Italy}
                
\date{\today}

%%%%%%%%%%%%%%%%%%%%%%%%%%%%%%%%%%%%%%%%%%%%%%%%%%%%%%%%%%%%%%%%%%%%%%%%%%%%%%%%%%%%%%%%%%%%%%%%%%%%
%%%%%%%%%%%%%%%%%%%%%%%%%%%%%%%%%%%%%%%%%%%%%%%%%%%%%%%%%%%%%%%%%%%%%%%%%%%%%%%%%%%%%%%%%%%%%%%%%%%%
\begin{abstract}

Many natural phenomena exhibit a stochastic nature that one attempts at modeling by using stochastic processes of different types. In this context, often one is interested in investigating the memory properties of the natural phenomenon at hand.  This is usually accomplished by computing the autocorrelation function of the numerical series describing the considered phenomenon. Often, especially when considering real world data, the autocorrelation function must be computed starting from a single numerical series: i.e. with a time-average approach.

Hereafter, we will propose a novel way of evaluating the time-average autocorrelation function, based on the preliminary evaluation of the quantity ${\cal{N}}(\tau,g_\mu,g_\nu)$, that, apart from normalization factors, represents a numerical estimate, based on a single realization of the process, of the 2-point joint probability density function $P(x_2,\tau; x_1,0)$. The main advantage of the proposed method is that it allows to quantitatively assess what is the {\em{error}} that one makes when numerically evaluating the autocorrelation function due to the fact that any simulated time series is necessarily bounded. In fact, we show that, for a wide class of stochastic processes admitting a nonlinear Langevin equation with white noise and that can be described by using a Fokker-Planck equation, the way the numerical estimate of the autocorrelation function converges to its theoretical prediction depends on the pdf tails. Moreover, the knowledge of ${\cal{N}}(\tau,g_\mu,g_\nu)$ allows to easily compute the process histogram and to characterize processes with multiple timescales. 

We will show the effectiveness of our new methodology by considering three stochastic processes whose autocorrelation function and two-point probability density function are both known in an analytical or numerical form, thus allowing direct comparisons.

\end{abstract}

\maketitle

%%%%%%%%%%%%%%%%%%%%%%%%%%%%%%%%%%%%%%%%%%%%%%%%%%%%%%%%%%%%%%%%%%%%%%%%%%%%%%%%%%%%%%%%%%%%%%%%%%%%
%%%%%%%%%%%%%%%%%%%%%%%%%%%%%%%%%%%%%%%%%%%%%%%%%%%%%%%%%%%%%%%%%%%%%%%%%%%%%%%%%%%%%%%%%%%%%%%%%%%%
\section{Introduction} \label{intro}
%%%%%%%%%%%%%%%%%%%%%%%%%%%%%%%%%%%%%%%%%%%%%%%%%%%%%%%%%%%%%%%%%%%%%%%%%%%%%%%%%%%%%%%%%%%%%%%%%%%%
%%%%%%%%%%%%%%%%%%%%%%%%%%%%%%%%%%%%%%%%%%%%%%%%%%%%%%%%%%%%%%%%%%%%%%%%%%%%%%%%%%%%%%%%%%%%%%%%%%%%

Stochastic processes are ubiquitous in different and heterogeneous research fields such as physics \cite{BM,VanKampen81,risken,gardiner, Schuss,Oksendal}, genomics \cite{Waterman,Durbin}, finance \cite{Bouchaud,Mantegna}, climatology \cite{vanStorch} and social sciences \cite{Helbing}.  In fact, many natural phenomena exhibit a stochastic nature that one attempts at modeling by using stochastic processes of different types \cite{beran, embrechts, Hull, BS1973, HW1987, Heston1993,archgarch,ctrw,levy}.

A relevant aspect one is usually interested in, regards the possibility of investigating the memory properties of the natural phenomenon at hand. In the context of stochastic processes this is usually accomplished by computing the autocorrelation function of the numerical series describing the considered phenomenon. Other tools certainly exist, such as for example the mean square displacement \cite{msd1, msd2} or the spectral function \cite{wk1,wk2,wk3}. But certainly the autocorrelation function plays a central role.

In many cases, the evaluation of the autocorrelation function is not a trivial task. Two different situations are indeed possible. In a first case one has to do with an empirical time series and the underlying stochastic model is not known. When this happens one must rely only on the computation of the empirical autocorrelation function, starting from a single available numerical series. The second case is when a model is available. In simple cases the existence of a model allows an analytical evaluation of the autocorrelation function. However in the vast majority of cases this is not possible and the evaluation of the autocorrelation function must be done, as for the previous case, starting from a surrogate time series that is obtained through numerical simulations of the considered model. The advantage, here, is that one can generate many numerical series.

One possible way for obtaining the autocorrelation function is to perform a time-average calculation. The other alternative is that of performing ensemble average simulations, when it is possible to generate many numerical series of the same process. Hereafter, we will consider the case of a time-average evaluation of the autocorrelation function, as it covers the case of an empirical numerical series. We will propose a novel way of evaluating the time-average autocorrelation function based on the computation of ${\cal{N}}(\tau,g_\mu,g_\nu)$, that gives the number of process outcomes that are in the bin $[g_{\mu-1}, g_\mu]$ at a certain a certain step $i$ and are in the bin $[g_{\nu-1}, g_\nu]$ at the step $(i + \tau)$, as we will see hereafter. The central result of our work is that the autocovariance $R(\tau)$ can be written as $R(\tau)= \sum_{\mu=1}^n C(\tau,\mu)$, where $C(\tau,\mu)$ is obtained from ${\cal{N}}(\tau,g_\mu,g_\nu)$ by averaging over the $\nu$-th bin. The main advantage of the proposed method is that the knowledge of $C(\tau,\mu)$ allows to understand which parts of the numerical series most contribute to the autocorrelation function at each lag. In fact we show that for a wide class of stochastic processes admitting a nonlinear Langevin equation with white noise and that can be described by using a Fokker-Planck equation, the way the numerical estimate of the autocorrelation function converges to its theoretical prediction depends on the pdf tails. Furthermore, our protocol allows to easily compute important quantities such as the process histogram or the conditional entropy.

Our methodology is also useful for characterizing processes with multiple time-scales. In fact, we show that when the process is characterized by the presence of one single dominating timescale, i.e. when the autocorrelation tail is well described by an effective exponential function, then $C(\tau,\mu)$ can be be factorized into $C(\tau,\mu) \approx F(x) {\cal{R}}(\tau)$ where numerical evidences show that  ${\cal{R}}(\tau)\approx R(\tau)$. In the case when the stochastic process is a truly multiscale one, i.e. its autocorrelation function can not be approximated by an exponential function, then the above factorization is no longer observed.
%Finally, our approach sets a bridge between the time-average computation and the ensemble average computation of the autocovariance function. 

We will show the effectiveness of our new methodology by considering a few well known stochastic processes whose autocorrelation function and two-point probability density function are both known in an analytical or numerical form, thus allowing direct comparisons.

The paper is organized as follows: in section \ref{model} we will illustrate the novel way of evaluating the autocorrelation function. Afterwards, in section \ref{OU} we will show the correctness of our method by reproducing the autocorrelation function of the Ornstein-Uhlembeck process. In section \ref{SW} and section \ref{RISK} we will do the same for two peculiar multi-scale stochastic processes. Our conclusions will be presented in section \ref{concl}.
 
%%%%%%%%%%%%%%%%%%%%%%%%%%%%%%%%%%%%%%%%%%%%%%%%%%%%%%%%%%%%%%%%%%%%%%%%%%%%%%%%%%%%%%%%%%%%%%%%%%%%
%%%%%%%%%%%%%%%%%%%%%%%%%%%%%%%%%%%%%%%%%%%%%%%%%%%%%%%%%%%%%%%%%%%%%%%%%%%%%%%%%%%%%%%%%%%%%%%%%%%%
\section{Methodology} \label{model}
%%%%%%%%%%%%%%%%%%%%%%%%%%%%%%%%%%%%%%%%%%%%%%%%%%%%%%%%%%%%%%%%%%%%%%%%%%%%%%%%%%%%%%%%%%%%%%%%%%%%
%%%%%%%%%%%%%%%%%%%%%%%%%%%%%%%%%%%%%%%%%%%%%%%%%%%%%%%%%%%%%%%%%%%%%%%%%%%%%%%%%%%%%%%%%%%%%%%%%%%%

Given one single empirical or simulated stationary numerical series $X=\{x_1, x_2, \cdots , x_T\}$ of length $T$, the time-average evaluation of the autocorrelation function is given by 
\begin{eqnarray}
                          &&  \rho_{TA}(\tau) = \frac{R_{TA}(\tau) - m \, m_\tau}{\sqrt{v \, v_\tau}} \qquad \qquad \quad
                                 R_{TA}(\tau)=\overline{X(0) \, X(\tau)} = \frac{1}{T-\tau_M} \, \sum_{i=1}^{T-\tau_M} \, x_i \, x_{i+\tau} \nonumber \\
                          && m=\frac{1}{T-\tau_M} \, \sum_{i=1}^{T-\tau_M} \, x_i \qquad \qquad  \qquad
                                m_\tau=\frac{1}{T-\tau_M} \, \sum_{i=1}^{T-\tau_M} x_{i + \tau} \nonumber \\
                          && v=\frac{1}{T-\tau_M} \, \sum_{i=1}^{T-\tau_M} \, x_i^2 - m^2 \qquad \qquad 
                                v_\tau=\frac{1}{T-\tau_M} \, \sum_{i=1}^{T-\tau_M} x_{i + \tau}^2 - m_\tau^2    \label{ACtime}
\end{eqnarray}
where $\tau$ is an integer lag number, whose maximal value $\tau_M$ is usually much smaller than $T$.              

Let us now classify our $x_i$ values into a certain number of bins. To this end, let us call $x_m={\rm{min}}(X)$ and $x_M={\rm{max}}(X)$ the minimum and maximal value of the numerical series $X$ and let us now generate a grid $G=\{g_0, g_1, g_2, \cdots , g_n\}$ where $n$ is the number of the grid bins. Each bin has length $\Delta=(x_M-x_m)/n$. Naturally: $g_0=x_m$ and $g_n=x_M$ with $g_i=x_m+i \, \Delta$. The choice of $n$ is arbitrary. However, experienced researchers understand how to choose $n$ in relation to $T$. This grid is for example routinely used for the computation of the histogram of the considered numerical series $X$. The requirement that all bins have the same length is not compulsory. However, this is the simplest choice.

Let us now consider the quantity ${\cal{N}}(\tau,g_\mu,g_\nu)$ that gives the number of elements of $X$ that are in the bin $[g_{\mu-1}, g_\mu]$ at a certain a certain step $i$ and are in the bin $[g_{\nu-1}, g_\nu]$ at the step $(i + \tau)$. Since we assume stationarity, we disregard the temporal dependance from the $i$-th step.

We claim that:
\begin{eqnarray}
                          && R_{TA}(\tau)=\frac{1}{T} \sum_{\mu=1}^n  \Bigl( \frac{g_{\mu-1}+g_\mu}{2}\Bigl)   \, 
                                                              \sum_{\nu=1}^n  \Bigl( \frac{g_{\nu-1}+g_\nu}{2}\Bigl) \, 
                                                              {\cal{N}}(\tau,g_\mu,g_\nu) \label{ACnew}
\end{eqnarray}
This is the central result of our work. We will illustrate the correctness of our claim in the next three sections by considering three simple stochastic processes. 
%It is worth recalling that if all bins in the grid $G$ have the same length then $|g_\mu - g_{\mu-1}|=\Delta$. 
Moreover, we define:
\begin{eqnarray}
                          && C(\tau,\mu)=\frac{1}{T}  \Bigl( \frac{g_{\mu-1}+g_\mu}{2}\Bigl)   \, 
                                                              \sum_{\nu=1}^n  \Bigl( \frac{g_{\nu-1}+g_\nu}{2}\Bigl) \, 
                                                              {\cal{N}}(\tau,g_\mu,g_\nu) \label{ACcontribdef} \\
                          && R_{TA}(\tau)= \sum_{\mu=1}^n C(\tau,\mu)  \nonumber
\end{eqnarray}
The quantity $C(\tau,\mu)$ allows to estimate the contribution of each bin to the autocorrelation function for each lag. In particular, we will show in the next sections that $C(\tau,\mu)$ allows us to quantitatively understand where the main contributions to $R(\tau)$ come from and to assess what is the {\em{error}} that one makes when numerically evaluating the autocorrelation function due to the fact that any simulated time series is bounded, i.e. it can never extend to infinity.

Finally, it is worth mentioning that the histogram of the considered numerical series, i.e. the number of elements of $X$ present in the $\mu$-th bin $[g_{\mu-1}, g_\mu]$, can be easily obtained by the following relation:
\begin{eqnarray}
                          && H(\mu)=\frac{1}{\tau_M}  \sum_{\tau=1}^{\tau_M} \sum_{\nu=1}^n  \,{\cal{N}}(\tau,g_\mu,g_\nu) \label{ISTOdef} 
\end{eqnarray}

The core of our methodology is the computation of ${\cal{N}}(\tau,g_\mu,g_\nu)$. Apart from normalization factors, as we will see hereafter this quantity provides a numerical estimate, based on a single realization of the process,  for the 2-point joint probability density function $P(x_2,\tau; x_1,0)$. The latter is needed when estimating the autocovariance function using the information about the stochastic process probability density functions:
\begin{eqnarray}
                           R_{EA}(\tau) = \int_{-\infty}^{+\infty} dx_2 \int_{-\infty}^{+\infty} d x_1 \, x_2 \, x_1 \, P(x_2,\tau; x_1,0) \label{ACprob}
\end{eqnarray}
as it is the case when considering ensemble-average simulations.

%%%%%%%%%%%%%%%%%%%%%%%%%%%%%%%%%%%%%%%%%%%%%%%%%%%%%%%%%%%%%%%%%%%%%%%%%%%%%%%%%%%%%%%%%%%%%%%%%%%%
%%%%%%%%%%%%%%%%%%%%%%%%%%%%%%%%%%%%%%%%%%%%%%%%%%%%%%%%%%%%%%%%%%%%%%%%%%%%%%%%%%%%%%%%%%%%%%%%%%%%
\section{Ornstein-Uhlembeck process} \label{OU}
%%%%%%%%%%%%%%%%%%%%%%%%%%%%%%%%%%%%%%%%%%%%%%%%%%%%%%%%%%%%%%%%%%%%%%%%%%%%%%%%%%%%%%%%%%%%%%%%%%%%
%%%%%%%%%%%%%%%%%%%%%%%%%%%%%%%%%%%%%%%%%%%%%%%%%%%%%%%%%%%%%%%%%%%%%%%%%%%%%%%%%%%%%%%%%%%%%%%%%%%%

Let us consider the Ornstein-Uhlembeck (OU) process \cite{ornstein} defined by the following Langevin equation:
\begin{eqnarray}
          &&  \dot{x}(t)=-\gamma \, x(t)+ D \, \Gamma(t)   \label{D1OU}                 
\end{eqnarray}
where $\Gamma(t)$ is a $\delta$--correlated Gaussian noise term with null average and unitary variance.  The OU 2-points probability density function is given by \cite{risken}: 
\begin{eqnarray}
            P^{(OU)}(x_2,t_2; x_1,t_1)={ \gamma \over 2 \pi D^2} \, 
                                             { 1 \over \sqrt{1-e^{- 2 \gamma \tau}}}~ 
                                             {\rm{exp}}\Bigl(-{ \gamma \over{2 D (1-e^{- 2 \gamma \tau})}}(x_1^2 + x_2^2 - 2 \, x_1 \, x_2 \, e^{-\gamma \tau} )
                                                            \Bigr)    \label{OUCongProb}
\end{eqnarray}
where $\tau=|t_2-t_1|$. The two parameters $D$ and $\gamma$ are related to the variance $v$ of the OU process by the relation $v=D^2/\gamma$. Hereafter we will assume $t_1=0$ due to stationarity and $D=1$ for the sake of simplicity.

Starting from Eq. \ref{OUCongProb}, it is easy to show that:
\begin{eqnarray}
                           R^{(OU)}(\tau) = \int_{-\infty}^{+\infty} dx_2 \int_{-\infty}^{+\infty} dx_1 \, x_1 \, x_2 \, P^{(OU)}(x_2,\tau; x_1,0)={1 \over  \gamma} \, e^{- \gamma \tau}  \label{ACOUana}
\end{eqnarray}

On the other hand, starting from Eq. \ref{D1OU} one can simulate a numerical series of the OU process and compute the autocorrelation function according to either Eq. \ref{ACtime} and Eq. \ref{ACnew}. In fact, we generated a time series of length $T=10^6$ with $\gamma=0.1$. Time discretization has been done with a time-step of $\Delta t=0.01$, i.e. we generated $100$ points per unit time and therefore a total number of $N=T/\Delta t=10^8$ process outcomes. Eq. \ref{D1OU} has been numerically integrated by using a simple Euler scheme \cite{mannella, platen}:
\begin{eqnarray}
                             x_i=x_{i-1}-\gamma \, x_{i-1} \, \Delta t + Z_i
\end{eqnarray}
where $x_0=0.1$ and $Z_i$ is randomly extracted from a normal distribution with zero average and variance $2 \, \Delta t$. In Table \ref{tab:ACOU} we show the autocorrelation function values of the OU process obtained at different lags (first column) by using the analytical expression of Eq. \ref{ACOUana} (second column), the time-average definition of Eq. \ref{ACtime} (third column) and the new computing method of Eq. \ref{ACnew} (fourth column), with $n=100$ and $\tau_M=50$. The values shown in the table refer to a single realization of the process with $x_m=-14.714$, $x_M=15.842$ and therefore $\Delta=0.30$. The table shows that there is a very good agreement between the two ways of evaluating the autocorrelation function given by Eq. \ref{ACtime} and Eq. \ref{ACnew}.
\begin{table} [H]
\small
\caption{OU process. Autocorrelation function values of the OU process obtained at different lags (first column) by using the analytical expression of Eq. \ref{ACOUana} (second column), the time-average definition of Eq. \ref{ACtime} (third column) and the new computing method of Eq. \ref{ACnew} (fourth column), with $n=100$ and $\tau_M=50$. The values shown in the table refer to a single realization of the process with $x_m=-14.714$, $x_M=15.842$ and therefore $\Delta=0.30$.}
\begin{tabular}{cccc}
\textbf{lag}	& \textbf{analytical}	& \textbf{time-average} & \textbf{new}	\\
 1 & 9.04837 & 9.06551 & 9.06461 \\
 2 & 8.18731 & 8.20402 & 8.20316 \\
 3 & 7.40818 & 7.42425 & 7.42354 \\
 4 & 6.7032 & 6.71676 & 6.71609 \\
 5 & 6.06531 & 6.07728 & 6.07693 \\
 6 & 5.48812 & 5.50068 & 5.50057 \\
 7 & 4.96585 & 4.97834 & 4.97833 \\
 8 & 4.49329 & 4.50509 & 4.5052 \\
 9 & 4.0657 & 4.08213 & 4.08219 \\
 10 & 3.67879 & 3.69969 & 3.69973 \\
 11 & 3.32871 & 3.35195 & 3.35178 \\
 12 & 3.01194 & 3.03627 & 3.03581 \\
 13 & 2.72532 & 2.7507 & 2.75029 \\
 14 & 2.46597 & 2.49335 & 2.49265 \\
 15 & 2.2313 & 2.2594 & 2.25891 \\
 16 & 2.01897 & 2.04964 & 2.04909 \\
 17 & 1.82684 & 1.85999 & 1.85934 \\
 18 & 1.65299 & 1.68753 & 1.68701 \\
 19 & 1.49569 & 1.52843 & 1.52774 \\
 20 & 1.35335 & 1.38549 & 1.385 \\
 21 & 1.22456 & 1.25669 & 1.25623 \\
 22 & 1.10803 & 1.14044 & 1.13981 \\
 23 & 1.00259 & 1.03904 & 1.03825 \\
 24 & 0.90718 & 0.94556 & 0.94497 \\
 25 & 0.82085 & 0.861093 & 0.860438 \\
 26 & 0.742736 & 0.786386 & 0.785509 \\
 27 & 0.672055 & 0.716723 & 0.715778 \\
 28 & 0.608101 & 0.651964 & 0.651006 \\
 29 & 0.550232 & 0.593744 & 0.592832 \\
 30 & 0.497871 & 0.541256 & 0.540313 \\
 31 & 0.450492 & 0.492942 & 0.492113 \\
 32 & 0.407622 & 0.450351 & 0.449758 \\
 33 & 0.368832 & 0.412915 & 0.412507 \\
 34 & 0.333733 & 0.378066 & 0.377765 \\
 35 & 0.301974 & 0.341923 & 0.341429 \\
 36 & 0.273237 & 0.307654 & 0.30737 \\
 37 & 0.247235 & 0.276972 & 0.276836 \\
 38 & 0.223708 & 0.248293 & 0.248356 \\
 39 & 0.202419 & 0.224896 & 0.224988 \\
 40 & 0.183156 & 0.204595 & 0.204728 \\
 41 & 0.165727 & 0.185518 & 0.185739 \\
 42 & 0.149956 & 0.168039 & 0.167781 \\
 43 & 0.135686 & 0.151847 & 0.151608 \\
 44 & 0.122773 & 0.137162 & 0.136886 \\
 45 & 0.11109 & 0.123923 & 0.123715 \\
 46 & 0.100518 & 0.109588 & 0.109639 \\
 47 & 0.0909528 & 0.100298 & 0.100412 \\
 48 & 0.0822975 & 0.0892468 & 0.089484 \\
 49 & 0.0744658 & 0.0797757 & 0.0801115 \\
 50 & 0.0673795 & 0.0708994 & 0.0712258 \\
\end{tabular}
\label{tab:ACOU}
\end{table}

\begin{figure}[H]
\begin{center}
\includegraphics[scale=0.65]{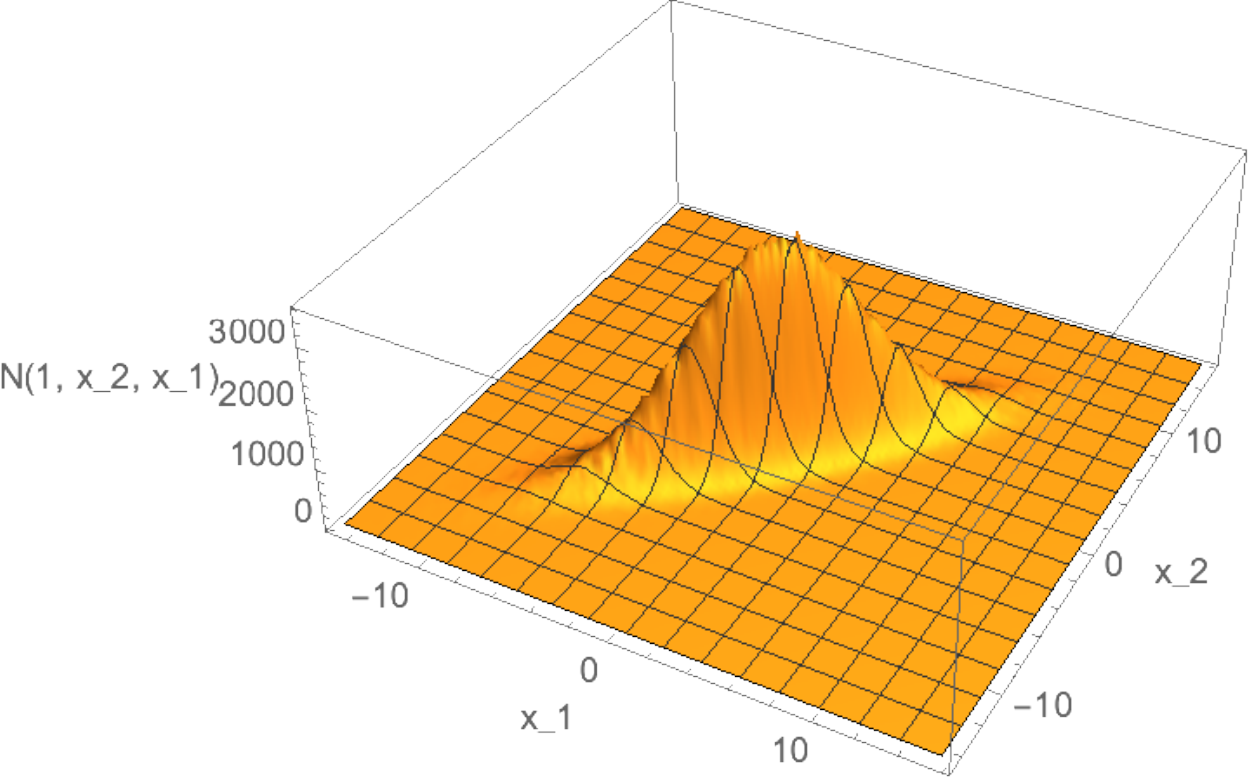}
\includegraphics[scale=0.74]{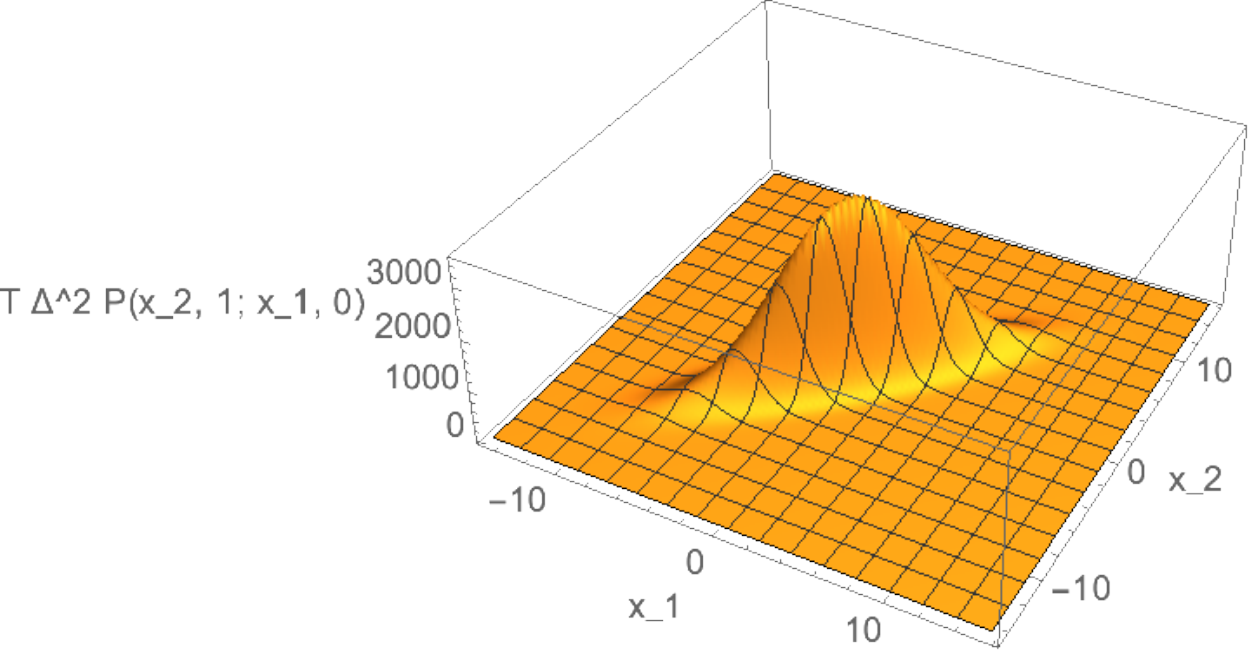}
\end{center}
\caption{OU process. Graphical 3-D representation of ${\cal{N}}(\tau,g_\mu,g_\nu)$ for $\tau=1$. In the left panel of Fig. \ref{ENNE_OU_tau1} we show ${\cal{N}}(\tau,g_\mu,g_\nu)$ for $\tau=1$. In the right panel of Fig. \ref{ENNE_OU_tau1} we show the quantity $P(x_2,\tau; x_1,0)  \,T \, \Delta^2$ whose analytical expression is given in Eq. \ref{OUCongProb}.}\label{ENNE_OU_tau1}
\end{figure}

In the left panel of Fig. \ref{ENNE_OU_tau1} we show ${\cal{N}}(\tau,g_\mu,g_\nu)$ for $\tau=1$. At fixed values of $\tau$, ${\cal{N}}(\tau,g_\mu,g_\nu)$ quantifies the joint occurrences in two different bins of our grid $G$. Therefore, ${\cal{N}}(\tau,g_\mu,g_\nu)$ is a proxy of the 2-point probability density function $P(x_2,\tau; x_1,0)$ whose analytical expression is given in Eq. \ref{OUCongProb}. In fact, in the right panel of Fig. \ref{ENNE_OU_tau1} we show the quantity $P(x_2,\tau; x_1,0)  \,T \, \Delta^2$, which is in agreement with the plot of the left panel. The normalization factor is $T \Delta^2$, which is the standard normalization factor to be used when going from occurrences to probability density functions in a bi-dimensional case.

\begin{figure}[H]
\begin{center}
{\includegraphics[scale=0.45]{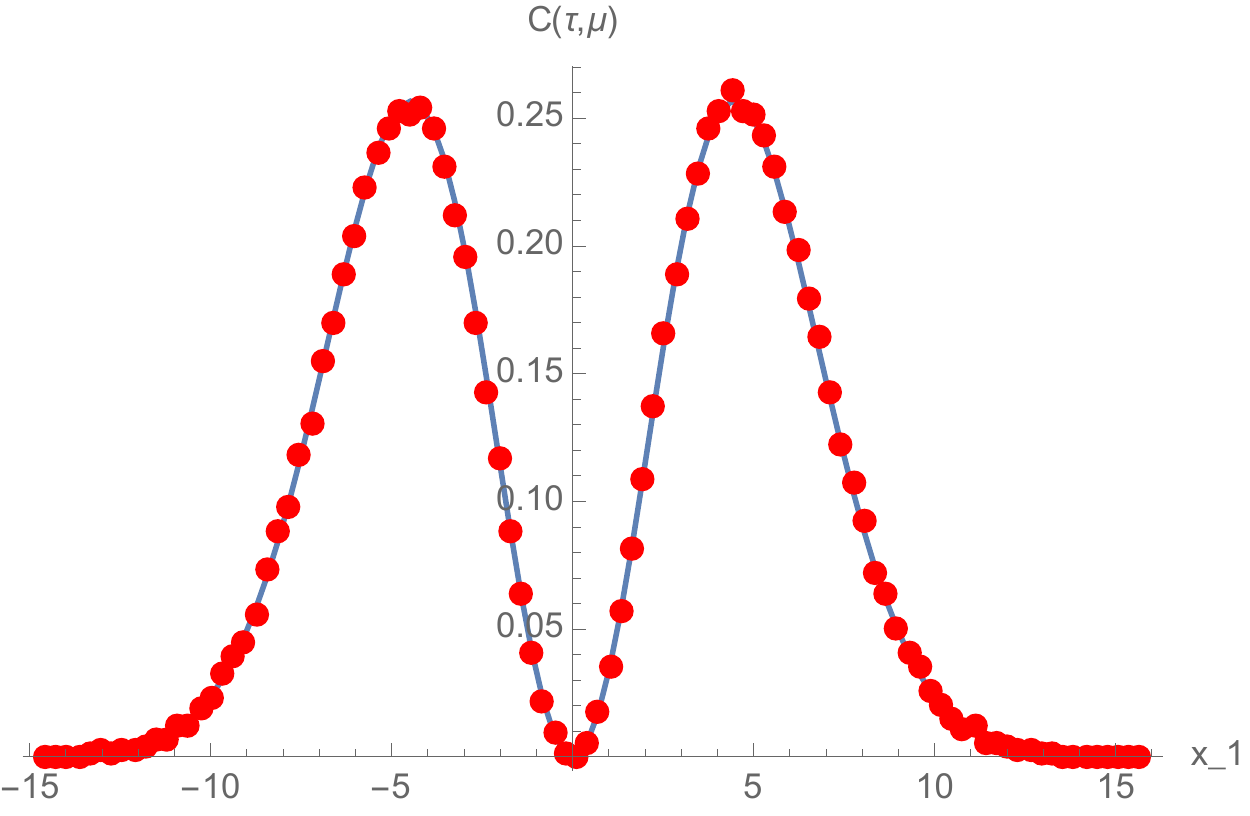}
 \includegraphics[scale=0.45]{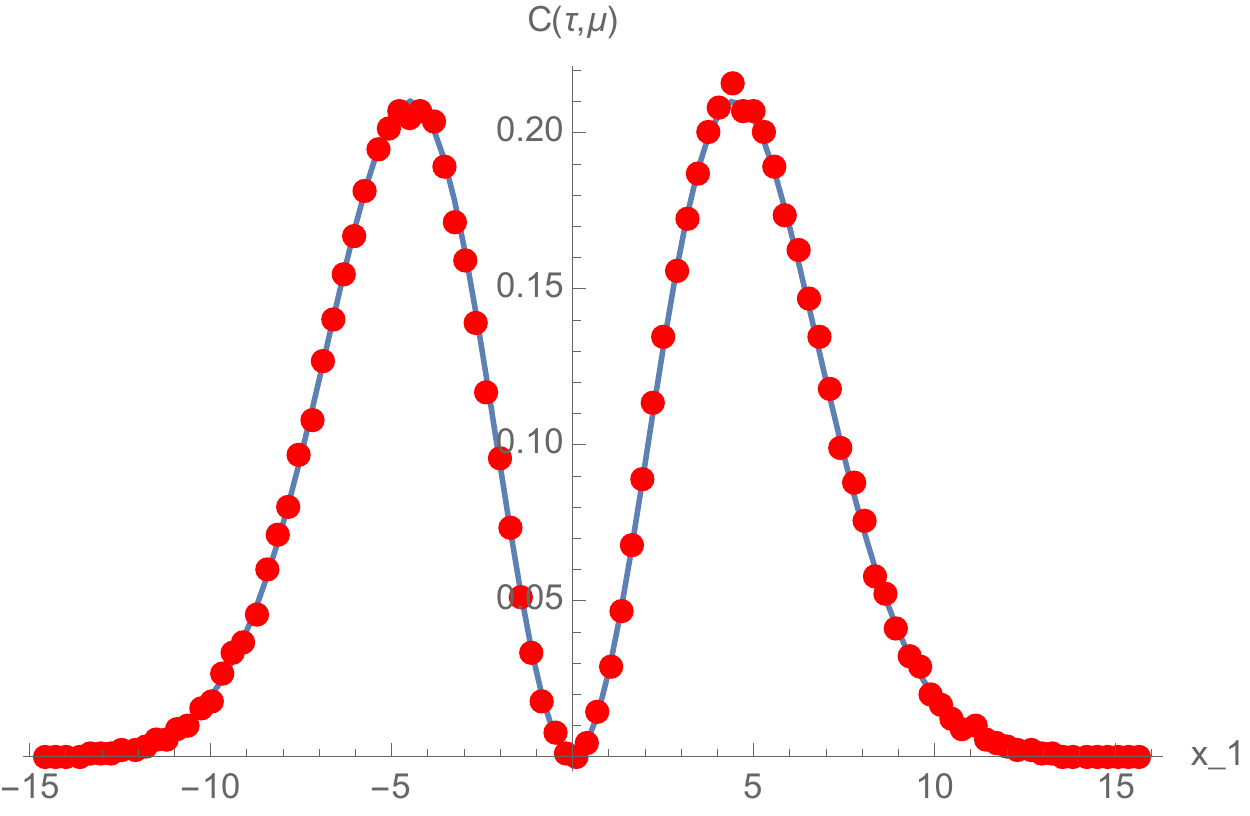}
 \includegraphics[scale=0.45]{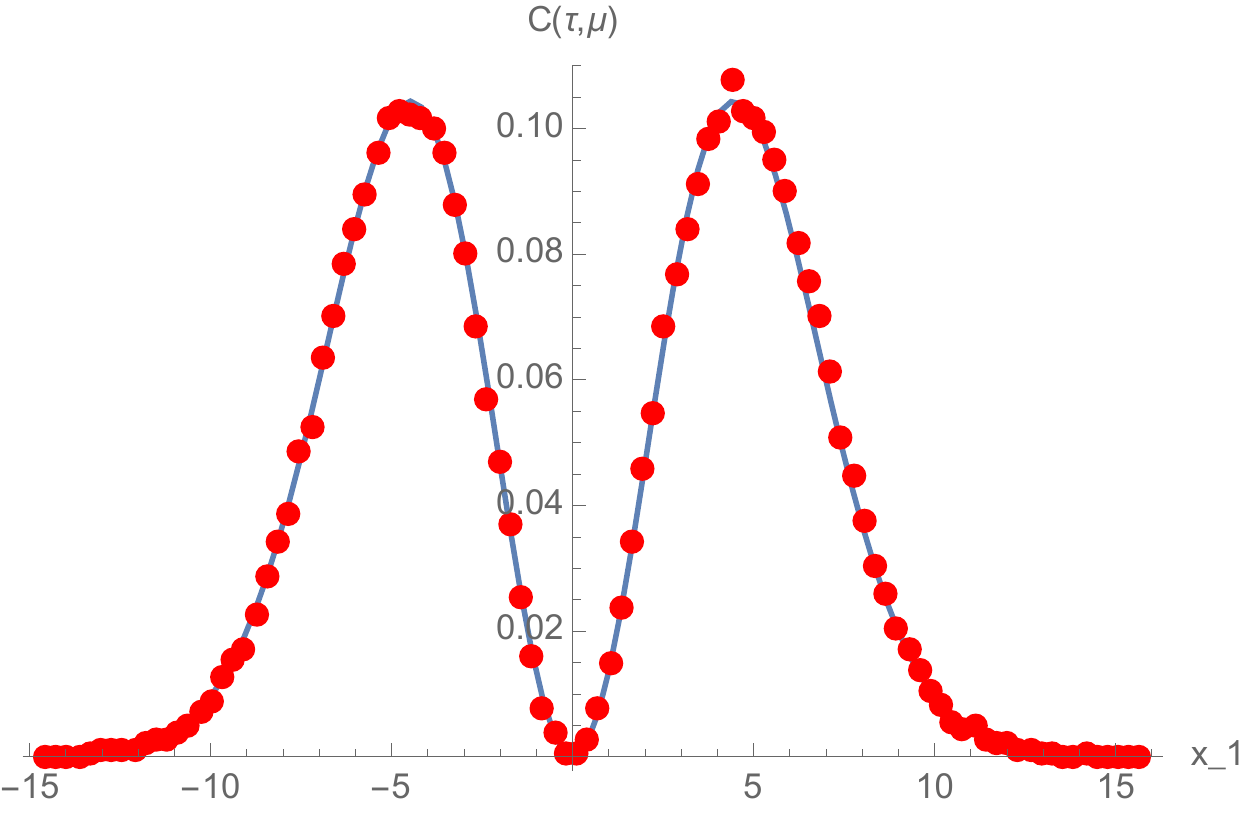}}\\
{\includegraphics[scale=0.45]{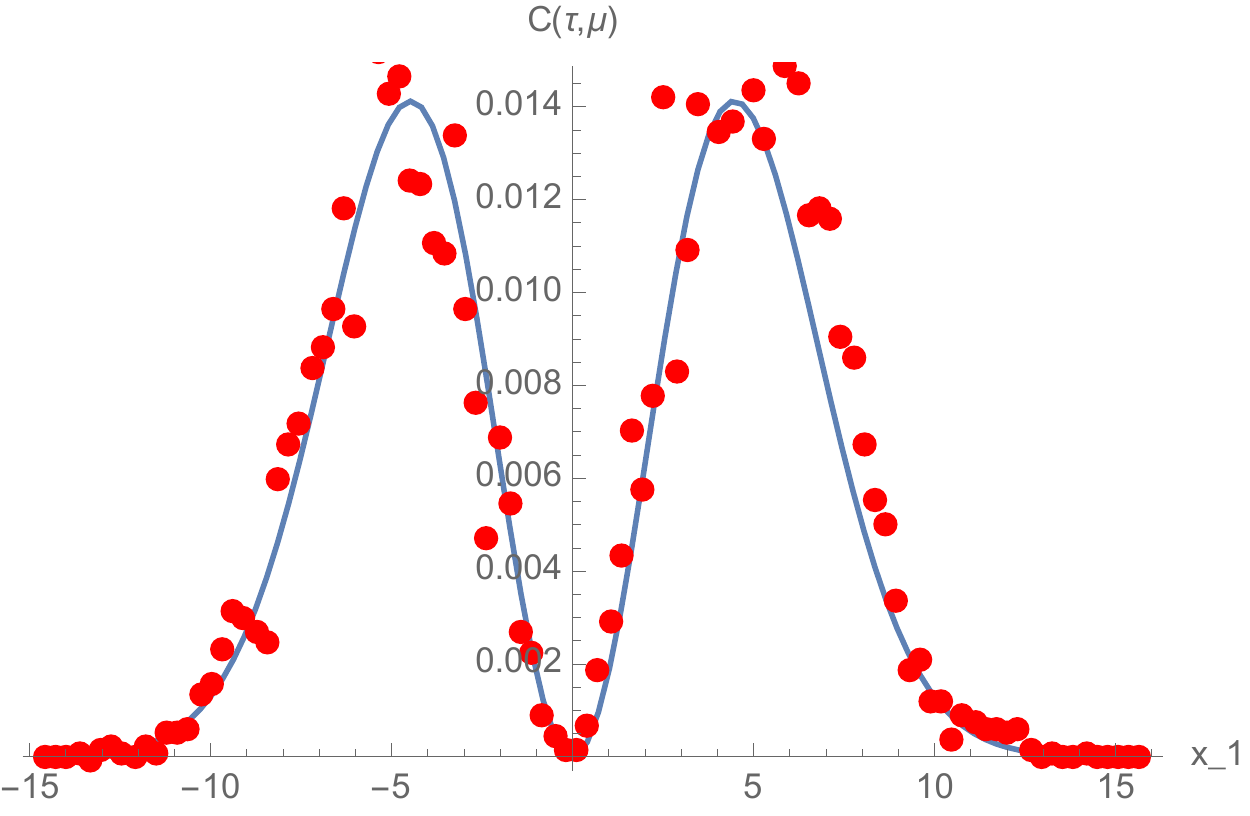}
 \includegraphics[scale=0.45]{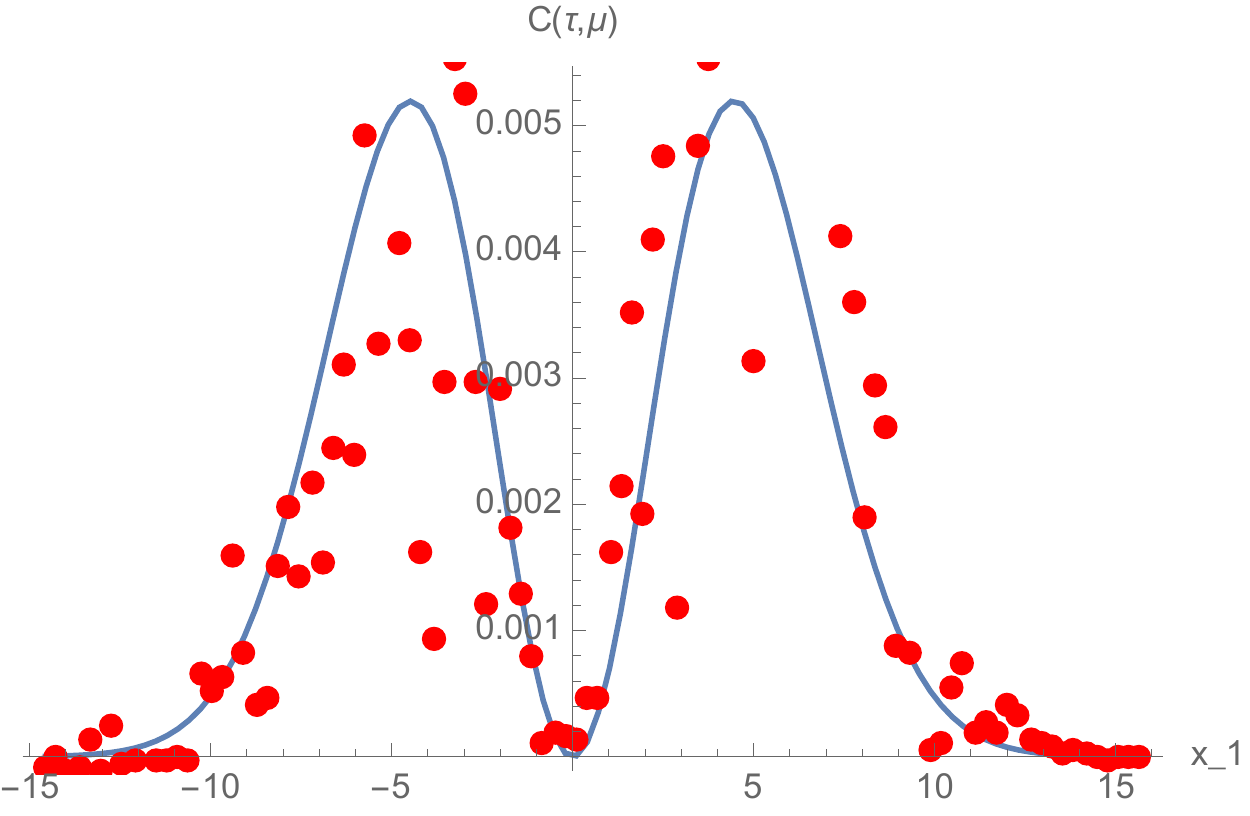}
 \includegraphics[scale=0.45]{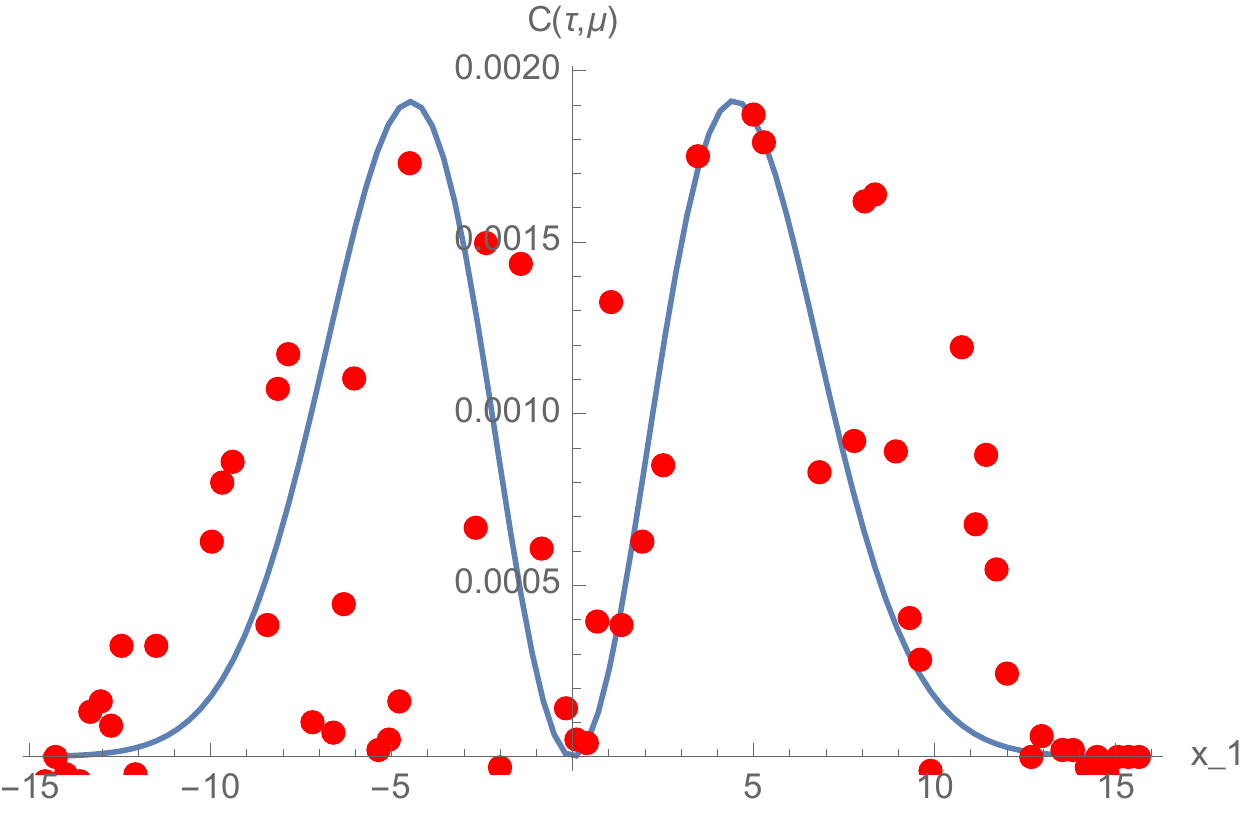}}
\end{center}
\caption{OU process. The red circles show the quantity $C(\tau,\mu)$ of Eq. \ref{ACcontribdef} for different lags. The top panels show the contributions at $\tau=1$, $\tau=3$, $\tau=10$ from left to right, while the bottom panels show the contributions at $\tau=30$, $\tau=40$, $\tau=50$ from left to right. The solid lines represent the theoretical predictions. They are obtained starting from Eq. \ref{ACcontribdef} where the empirically quantity  ${\cal{N}}(\tau,g_\mu,g_\nu)$ is replaced by its theoretical counterpart $P^{(OU)}(x_2,\tau; x_1,0)  \,T \, \Delta^2$ with $x_1=(g_{\mu-1}+g_\mu)/2$ and $x_2=(g_{\nu-1}+g_\nu)/2$}\label{contrib_OU}
\end{figure}

In the various panels of Fig. \ref{contrib_OU} the red circles show the quantity $C(\tau,\mu)$ of Eq. \ref{ACcontribdef} for different lags. The top panels show the contributions at $\tau=1$, $\tau=3$, $\tau=10$ from left to right, while the bottom panels show the contributions at $\tau=30$, $\tau=40$, $\tau=50$ from left to right. The solid lines represent the theoretical predictions. They are obtained starting from Eq. \ref{ACcontribdef} where the empirically quantity  ${\cal{N}}(\tau,g_\mu,g_\nu)$ is replaced by its theoretical counterpart $P^{(OU)}(x_2,\tau; x_1,0)  \,T \, \Delta^2$ with $x_1=(g_{\mu-1}+g_\mu)/2$ and $x_2=(g_{\nu-1}+g_\nu)/2$.

One can easily notice that the agreement between theoretical prediction and simulations degrades as long as $\tau$ increases. This corresponds to the well known fact that having reliable predictions for the autocorrelation function for large lags gets increasingly difficult. Usually, better estimates of $R(\tau)$ for large lag values are obtained by increasing $T$. A thorough discussion on this topic is out of the scope of the present work. We here want instead to emphasize the kind of information that $C(\tau,\mu)$ can give us. To this end let us consider that the theoretical counterpart of $C(\tau,\mu)$ is given by
\begin{eqnarray}
                           C_{th}^{(OU)}=\Delta^2 \, T \, \int_{-\infty}^{+\infty} \, dx_1\ \, x \, x_1 \, P(x, \tau; x_1, 0)= \Delta^2 \, T \, \sqrt{\frac{\gamma}{2 \, \pi}} \, x^2 \, e^{-\frac{1}{2} \gamma x^2} \, e^{- \gamma \tau}
\end{eqnarray}
This shows how the contributions $C^{(OU)}(\tau,\mu)$ are factorized into a form factor
\begin{eqnarray}
                           F^{(OU)}(x)= \Delta^2 \, T \, \sqrt{\frac{\gamma}{2 \, \pi}} \, x^2 \, e^{-\frac{1}{2} \gamma x^2}
\end{eqnarray}
that only depends on the spatial variable and a temporal part with an exponential cut-off $e^{- \gamma \tau}$: $C_{th}^{(OU)}=F^{(OU)}(x) \, e^{- \gamma \tau}$. This structure is also suggested by the panels of Fig. \ref{contrib_OU}. The specific shape of the form factor $F^{(OU)}(x)$ indicates that the contributions to the autocorrelation function do not come from values close to zero, but rather, from large x values, as expected. 

In order to quantify this effect let us now assume that  $C_{th}^{(OU)}$ has a cut-off for large x values. Let us therefore introduce $C_{th,L}^{(OU)}=C_{th}^{(OU)} \, H(L-x) \, H(L+x)$, where $H(\cdot)$ is the Heaviside function. In this way we can mimic what usually happens when simulating a stochastic process: due to the fact that $T$ is finite, we have a numerical series bounded by a certain maximal value $L$: $|x_i|\le L$. Therefore, simulations only allow us to estimate $C_{th,L}^{(OU)}$ rather the $C_{th}^{(OU)}$. We are therefore interested in understanding how the resulting autocorrelation function $R_L^{(OU)}(\tau)$ converges to $R^{(OU)}(\tau)$ as long as $L$ increases. By performing the relevant integrations we get:
\begin{eqnarray}
                           R_L^{(OU)}(\tau)=\frac{1}{\gamma} \, e^{- \gamma \tau} \, \Biggl(
                                                                                                                                 \frac{1}{\gamma} \, {\rm{Erf}}\Bigl( \frac{L \sqrt{\gamma}}{\sqrt{2}}\Bigl) - 
                                                                                                                                 \frac{\sqrt{2} L}{\sqrt{\pi \gamma}} e^{- \gamma/2 \, L^2  }
                                                                                                                        \Biggr)
\end{eqnarray}
This expression shows that $R_L^{(OU)}(\tau)$ converges to $R^{(OU)}(\tau)$ relatively fastly, due to the $e^{- \gamma/2 \, L^2}$ term. In fact, experienced researchers know that the OU autocorrelation function can be simulated in a quite simple way. The factor $e^{- \gamma/2 \, L^2}$ can be interpreted as an estimate of the {\em{error}} that one makes when numerically evaluating the autocorrelation function due to the fact that any simulated time series is necessarily bounded $|x_i|<L$.

%\begin{table}
%\small 
%\caption{computational time OU}
%\begin{tabular}{ccc}
%\textbf{$T$}	& \textbf{new} & \textbf{time-average}	\\
%$10^6$ & 0.778 & 0.267  \\
%$10^7$ & 8.333 & 2.637  \\
%$10^8$ & 60.643 & 29.263  \\
%$10^9$ & 577.163 & 284.157  \\
% \end{tabular}
%\label{tab:ACOUtime}
%\end{table}

%%%%%%%%%%%%%%%%%%%%%%%%%%%%%%%%%%%%%%%%%%%%%%%%%%%%%%%%%%%%%%%%%%%%%%%%%%%%%%%%%%%%%%%%%%%%%%%%%%%%
%%%%%%%%%%%%%%%%%%%%%%%%%%%%%%%%%%%%%%%%%%%%%%%%%%%%%%%%%%%%%%%%%%%%%%%%%%%%%%%%%%%%%%%%%%%%%%%%%%%%
\section{The Square Well process} \label{SW}
%%%%%%%%%%%%%%%%%%%%%%%%%%%%%%%%%%%%%%%%%%%%%%%%%%%%%%%%%%%%%%%%%%%%%%%%%%%%%%%%%%%%%%%%%%%%%%%%%%%%
%%%%%%%%%%%%%%%%%%%%%%%%%%%%%%%%%%%%%%%%%%%%%%%%%%%%%%%%%%%%%%%%%%%%%%%%%%%%%%%%%%%%%%%%%%%%%%%%%%%%

The second example we want to consider is given by the stochastic process described by a Langevin equation with additive noise and the following drift coefficient \cite{risken}:
\begin{eqnarray}
          &&  \dot{x}(t)=-h(x)+ D \, \Gamma(t)   \label{D1SW} \\
          &&  h(x) =\left    \lbrace             
          \begin{aligned}
	        &	\text{$0$ if $x \in [-L, +L]$} \\
	        &				\\
	        & 	\text{-$\frac{\pi}{L} \, \tan \Bigl( \frac{\pi}{2 L} \, x \Bigl)$  if $x \in [-L, +L]$} 
         \end{aligned}
         \right.
\end{eqnarray}
The reason for considering such process is that, by using the methodology of eigenfunction expansion \cite{risken,gardiner} it is possible to prove that the autocovariance function $R(\tau)$ of the above process is given by
\begin{eqnarray}
                          && R(\tau)=\sum_{n=0}^{\infty} \, c_n^2 e^{- \lambda_n \, \tau} \qquad \qquad \qquad 
                                c_n=\int_{-L}^{+L} \, dx \, x \, \psi_0(x) \, \psi_n(x) \label{ACSWana}\\ 
                          && \psi_0(x)=\frac{1}{L} \, \cos \Bigl( \frac{\pi}{2 L}\, x\Bigr) \qquad \qquad \quad
                                \lambda_0=0 \nonumber \\
                          && \psi_n(x)=\frac{1}{L} \, \cos \Bigl( \frac{\pi}{L}\, (n+\frac{1}{2}) \, x\Bigr)  \quad \quad
                                \lambda_n=\frac{\pi^2}{L^2}\,(n^2+n) \qquad \qquad 
                                n \, {\rm{even}} \nonumber \\
                          && \psi_n(x)=\frac{1}{L} \, \sin \Bigl( \frac{\pi}{L}\, n \, x\Bigr)  \qquad \quad \qquad
                                \lambda_n=\frac{\pi^2}{L^2}\,(n^2-\frac{1}{4}) \quad \quad \quad \, \, \, \, 
                                n \, {\rm{odd}}  \nonumber
\end{eqnarray}
The quantities $\{ \psi_n,\lambda_n \}$ are the eigenfunctions and eigenvalues of the Schr\"odinger equation with potential:
\begin{eqnarray}
                            V_S(x)=\frac{h(x)^2}{4}+\frac{1}{2}\, \frac {\partial h(x)}{\partial x}= \left \{ \begin{array}{cc}
                                - \frac{\pi^2}{4 \, L}   &{\rm{if}}~~|x| \leq L ,\\
                                     &   \\
                                \infty   &{\rm{if}}~~|x| > L .   
                        \end{array} \right.  \label{SWpot}
\end{eqnarray}
This potential describes a rectangular square well with infinite walls. Hereafter, we will refer to this process as the Square Well (SW) process. The result of Eq. \ref{ACSWana} implies that the SW process is characterized by the presence of an infinite set of discrete timescales, in contrast to the OU process which is characterized by the presence of just one single timescale $\gamma^{-1}$. The 2-point probability density function is given by \cite{risken,gardiner}
\begin{eqnarray}
                          && P^{SW}(x_2, \tau; x_1, 0)=\psi_0(x_2)^2 \, \psi_0(x_1)^2 + \sum_{n=1}^{\infty} \, \psi_0(x_2) \, \psi_n(x_2) \, \psi_0(x_1) \, \psi_n(x_1) \, e^{- \lambda_n \tau} \label{congprobthSW}
\end{eqnarray}

As much as in the previous case, we numerically integrated Eq. \ref{D1SW} by considering a simple Euler scheme with $x_0=0.1$ and $L=20$. In Table \ref{tab:ACSW} we show the autocorrelation function values of such process obtained at different lags (first column) by using the analytical expression of Eq. \ref{ACSWana} (second column), the time-average definition of Eq. \ref{ACtime} (third column) and the new computing method of Eq. \ref{ACnew} (fourth column), with $n=100$ and $\tau_M=50$. The values shown in the table refer to a single realization of the process with $x_m=-21.111$, $x_M=20.462$ and therefore $\Delta=0.416$. In obtaining the values of the second column we considered the first 100 terms in the sum only: unfortunately we were not able to obtain an analytical expression for $R(\tau)$ and we therefore have only a numerical estimation of the expected $R(\tau)$. The table shows a very good agreement between the two ways  of evaluating the autocorrelation function given by Eq. \ref{ACtime} and Eq. \ref{ACnew}.
\begin{table} [H]
\small
\caption{SW process. Autocorrelation function values of the SW process obtained at different lags (first column) by using the analytical expression of Eq. \ref{ACSWana} (second column), the time-average definition of Eq. \ref{ACtime} (third column) and the new computing method of Eq. \ref{ACnew} (fourth column), with $n=100$ and $\tau_M=50$. The values shown in the table refer to a single realization of the process with $x_m=-21.111$, $x_M=20.462$ and therefore $\Delta=0.416$. In obtaining the values of the second column we considered the first 100 terms in the sum only.}
\begin{tabular}{cccc}
\textbf{lag}	& \textbf{analytical}	& \textbf{time-average} & \textbf{new}	\\
 1 & 50.982 & 51.2245 & 51.2222 \\
 2 & 50.0434 & 50.2577 & 50.2556 \\
 3 & 49.1228 & 49.3136 & 49.3112 \\
 4 & 48.2197 & 48.3892 & 48.3869 \\
 5 & 47.3337 & 47.482 & 47.48 \\
 6 & 46.4643 & 46.595 & 46.5929 \\
 7 & 45.6111 & 45.7253 & 45.7231 \\
 8 & 44.7738 & 44.8754 & 44.8732 \\
 9 & 43.9521 & 44.0427 & 44.0405 \\
 10 & 43.1456 & 43.2269 & 43.225 \\
 11 & 42.354 & 42.4296 & 42.4281 \\
 12 & 41.577 & 41.6476 & 41.6461 \\
 13 & 40.8143 & 40.8845 & 40.8825 \\
 14 & 40.0657 & 40.1374 & 40.1352 \\
 15 & 39.3309 & 39.402 & 39.3998 \\
 16 & 38.6096 & 38.6819 & 38.6795 \\
 17 & 37.9015 & 37.974 & 37.9718 \\
 18 & 37.2065 & 37.2772 & 37.2749 \\
 19 & 36.5242 & 36.5914 & 36.5893 \\
 20 & 35.8544 & 35.9164 & 35.9143 \\
 21 & 35.197 & 35.2515 & 35.25 \\
 22 & 34.5516 & 34.5976 & 34.5962 \\
 23 & 33.918 & 33.9568 & 33.9555 \\
 24 & 33.2961 & 33.3266 & 33.3257 \\
 25 & 32.6856 & 32.7047 & 32.7037 \\
26 & 32.0862 & 32.0943 & 32.0936 \\
 27 & 31.4979 & 31.4944 & 31.4935 \\
 28 & 30.9204 & 30.9078 & 30.9071 \\
 29 & 30.3534 & 30.3331 & 30.3323 \\
 30 & 29.7969 & 29.7713 & 29.7705 \\
 31 & 29.2505 & 29.2196 & 29.2191 \\
 32 & 28.7142 & 28.6735 & 28.6733 \\
 33 & 28.1877 & 28.1361 & 28.1358 \\
 34 & 27.6709 & 27.6085 & 27.6081 \\
 35 & 27.1635 & 27.0943 & 27.0939 \\
 36 & 26.6655 & 26.5871 & 26.5867 \\
 37 & 26.1766 & 26.0905 & 26.0901 \\
 38 & 25.6966 & 25.6006 & 25.6005 \\
 39 & 25.2254 & 25.1188 & 25.1188 \\
 40 & 24.7629 & 24.6446 & 24.6447 \\
 41 & 24.3089 & 24.178 & 24.1782 \\
 42 & 23.8632 & 23.7209 & 23.721 \\
 43 & 23.4256 & 23.2703 & 23.2706 \\
 44 & 22.9961 & 22.8265 & 22.827 \\
 45 & 22.5745 & 22.3918 & 22.3923 \\
 46 & 22.1606 & 21.9632 & 21.9638 \\
 47 & 21.7542 & 21.544 & 21.5446 \\
 48 & 21.3554 & 21.1333 & 21.1339 \\
 49 & 20.9638 & 20.7306 & 20.7306 \\
 50 & 20.5794 & 20.3371 & 20.337 \\
\end{tabular}
\label{tab:ACSW}
\end{table}

The autocorrelation function that can be obtained form the numerical values reported in second column of Table \ref{tab:ACSW} can be fitted by an exponential function $e^{- \Lambda \tau}$ with $\Lambda=0.01879$. This result seems to be in contrast with the fact that this is a process with multiple timescales. However, since we have an infinite set of discrete timescales, it is evident that for large time lags only the largest timescale is relevant. The largest timescale in the process is that associates with the lowest value of $\lambda_n$. In the present case such lowest value is $\lambda_1=0.01850$, which is very close to the fitted value of $\Lambda$. %This simple argument shows that the process we are considering here is indeed similar to an OU process.

\begin{figure}[H]
\begin{center}
{\includegraphics[scale=0.65]{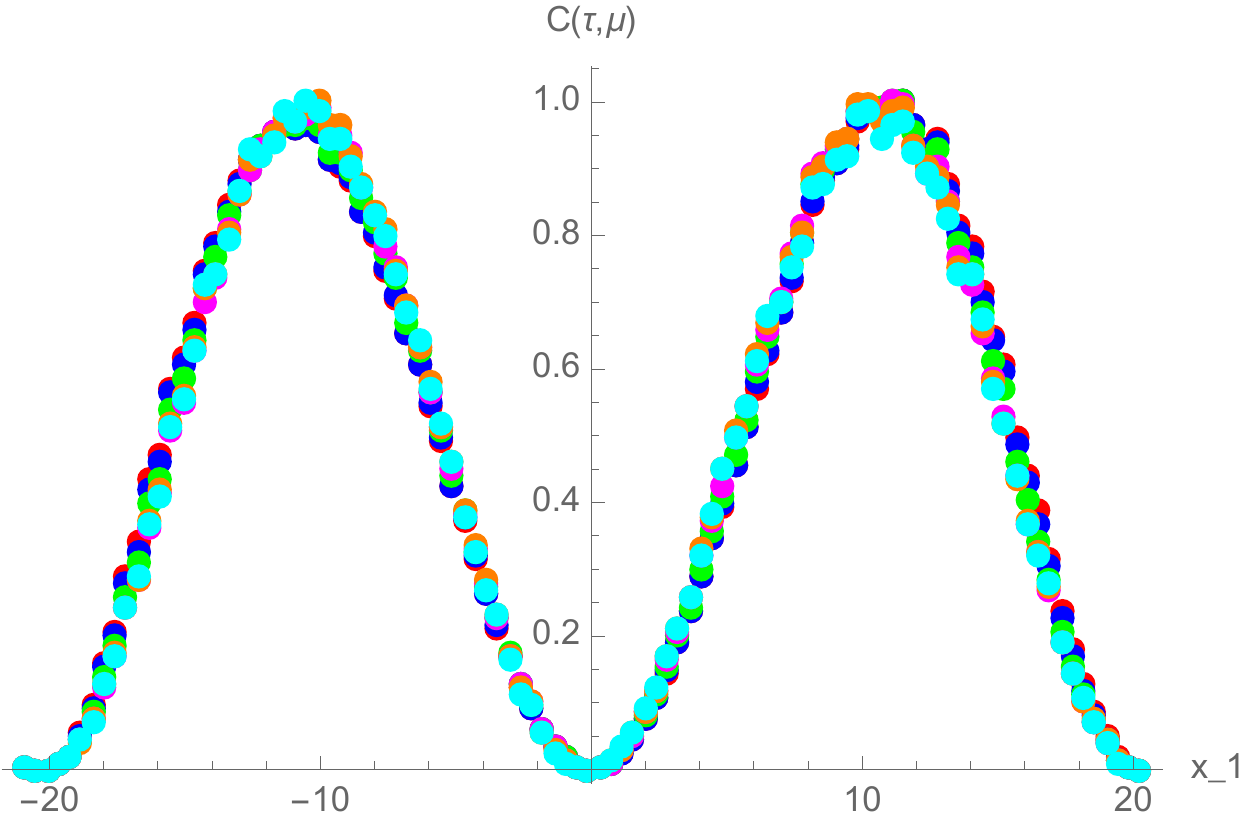}
 \includegraphics[scale=0.65]{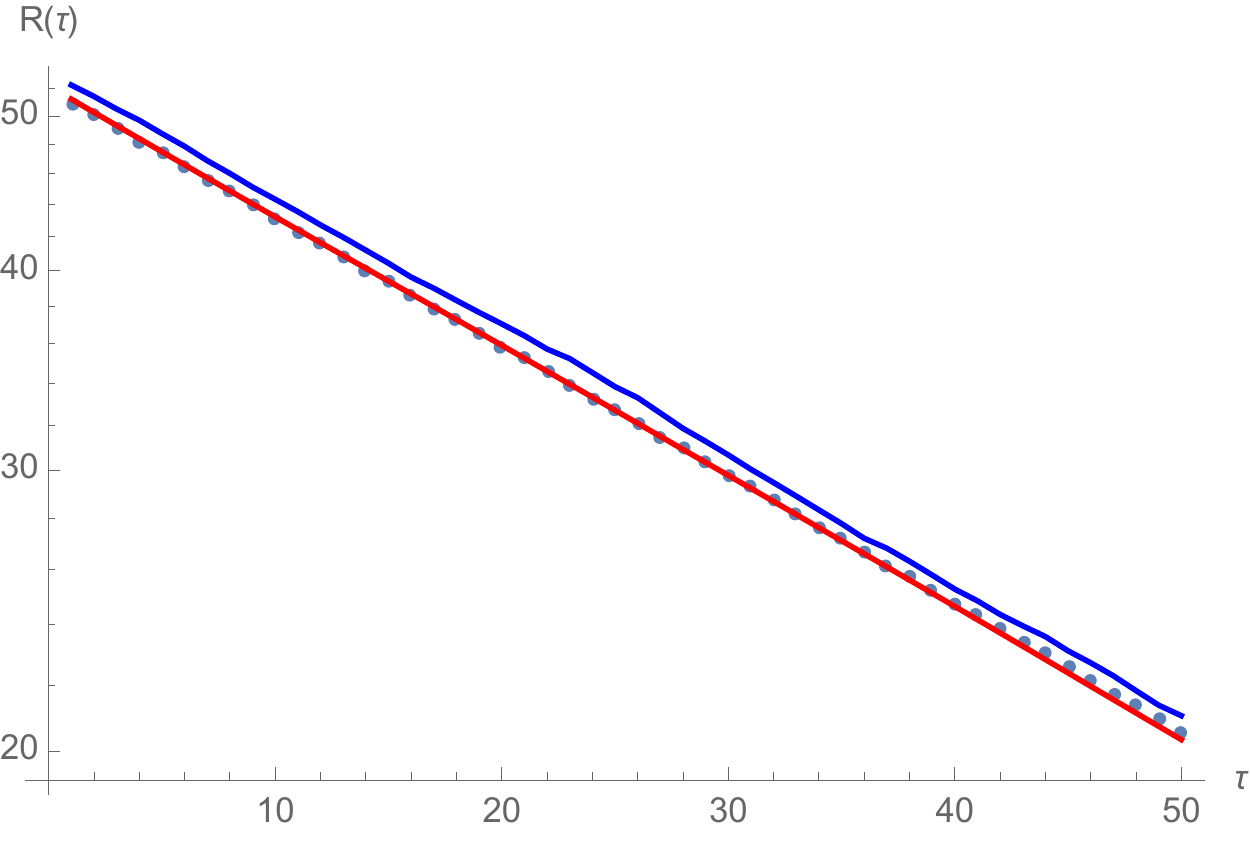}}
\end{center}
\caption{SW Process. The left panel shows the quantities $C^{(SW)}(\tau,\mu)$ of Eq. \ref{ACcontribdef} rescaled with their maximal value for different value of $\tau$: $\tau=1$ (red), $\tau=3$ (blue), $\tau=10$ (green), $\tau=30$ (magenta), $\tau=40$ (orange), $\tau=50$ (cyan). The blue line in the right panel of Fig.  \ref{contrib_SW} shows the maximal values of the $C^{(SW)}(\tau,\mu)$ curves shown in the left panel, i.e. the factors that have been used to normalize that curves. The red line in this panel corresponds to the values in the third column of Table \ref{tab:ACSW}, while the circles correspond to the theoretical prediction given by second column of Table \ref{tab:ACSW}. The right panel is in a log-linear scale.}\label{contrib_SW}
\end{figure}

In the left panel of Fig. \ref{contrib_SW} we show the quantities $C^{(SW)}(\tau,\mu)$ of Eq. \ref{ACcontribdef} rescaled with their maximal value for different value of $\tau$: $\tau=1$ (red), $\tau=3$ (blue), $\tau=10$ (green), $\tau=30$ (magenta), $\tau=40$ (orange), $\tau=50$ (cyan). It is evident that all these curves collapse onto a single one, thus indicating that also for this process the contributions $C^{(SW)}(\tau,\mu)$ are factorized into a form factor $F^{(SW)}(x)$ that only depends on the spatial variable and a temporal part with an exponential cut-off: $C^{(SW)}(\tau,\mu) = F^{(SW)}(x) {\cal{R}}^{(SW)}(\tau)$. The blue line in the right panel of Fig.  \ref{contrib_SW} shows the maximal values of the $C^{(SW)}(\tau,\mu)$ curves shown in the left panel, i.e. the factors that have been used to normalize those curves, i.e. the scaling factor ${\cal{R}}^{(SW)}(\tau)$. The red line in this panel corresponds to the values in the third column of Table \ref{tab:ACSW}, while the circles correspond to the theoretical prediction given by second column of Table \ref{tab:ACSW}. The results in this panel essentially shows that the scaling factor ${\cal{R}}^{(SW)}(\tau)$ is nothing but $R^{(SW)}(\tau)$. This is also confirmed when considering the theoretical counterparts of the contributions: 
\begin{eqnarray}
                           C_{th}^{(SW)}=\Delta^2 \, T \, \int_{-\infty}^{+\infty} \, dx_1 \, x \, x_1 \, P^{(SW)}(x, \tau; x_1, 0) \label{contribthSW}
\end{eqnarray}
Although we were not able to obtain $C_{th}^{(SW)}$ in a closed form, we can nevertheless compute it numerically. The fact that ${\cal{R}}^{(SW)}(\tau)=R^{(SW)}(\tau)$ is also observed in the case of the OU process where we have an exponential cut-off $e^{- \gamma \tau}$ that again corresponds to the OU autocorrelation function.
The shape of the form factor $F^{(SW)}(x)$ seems very similar to the one of the OU process, at least qualitatively. For this process we were not able to obtain $F^{(SW)}(x)$ in a closed form. 

%%%%%%%%%%%%%%%%%%%%%%%%%%%%%%%%%%%%%%%%%%%%%%%%%%%%%%%%%%%%%%%%%%%%%%%%%%%%%%%%%%%%%%%%%%%%%%%%%%%%
%%%%%%%%%%%%%%%%%%%%%%%%%%%%%%%%%%%%%%%%%%%%%%%%%%%%%%%%%%%%%%%%%%%%%%%%%%%%%%%%%%%%%%%%%%%%%%%%%%%%
\section{The DELTA process} \label{RISK}
%%%%%%%%%%%%%%%%%%%%%%%%%%%%%%%%%%%%%%%%%%%%%%%%%%%%%%%%%%%%%%%%%%%%%%%%%%%%%%%%%%%%%%%%%%%%%%%%%%%%
%%%%%%%%%%%%%%%%%%%%%%%%%%%%%%%%%%%%%%%%%%%%%%%%%%%%%%%%%%%%%%%%%%%%%%%%%%%%%%%%%%%%%%%%%%%%%%%%%%%%

Let us consider the stochastic process described by the following Langevin equation \cite{risken}:
\begin{eqnarray}
               && \dot{x}(t)=- h(x(t))\, + D\,\Gamma(t)  \nonumber \\
               &&  h(x)=\left \{ \begin{array}{cc}
                                +k   &{\rm{if}}~~x < 0 ,\\
                                     &   \\
                                -k   &{\rm{if}}~~x >  0 .   
                        \end{array} \right. \label{D1RISK}
\end{eqnarray}
where $k$ is a real constant and $\Gamma(t)$ is a $\delta$--correlated Gaussian noise term. By using the methodology of eigenfunction expansion \cite{risken,gardiner} it is possible to prove that the autocovariance function $R(\tau)$ of the above process is:
\begin{eqnarray}
                          && R(\tau)=\int_{k^2/4}^{\infty} \, c(E)^2 e^{- E \, \tau} \qquad \qquad \qquad 
                                c(E)=\int_{-\infty}^{+\infty} \, dx \, x \, \psi_0(x) \, \psi_E(x) \label{ACRISKana}\\ 
                          && \psi_0(x)=\sqrt{\frac{k}{2}} \, e^{- \frac{k}{2} |x|} \qquad \qquad \qquad  \qquad 
                                E=0 \nonumber \\
                          && \psi_E(x)=\frac{1}{\sqrt{2 \, \pi}} \, \frac{1}{(E - \frac{k^2}{4})^{1/4}} \, \cos \Bigl( \sqrt{E - \frac{k^2}{4}} \, x \Bigr) - 
                                                \frac{k}{2} \,  \frac{1}{\sqrt{2 \, \pi}} \, \frac{1}{(E - \frac{k^2}{4})^{3/4}} \, \sin \Bigl( \sqrt{E - \frac{k^2}{4}} \, |x| \Bigr)\quad 
                                E>\frac{k^2}{4} \qquad 
                                {\rm{even \, solution}} \nonumber \\
                          && \psi_E(x)=\frac{1}{\sqrt{2 \, \pi}} \, \frac{1}{(E - \frac{k^2}{4})^{1/4}} \, \sin \Bigl( \sqrt{E - \frac{k^2}{4}} \, x \Bigr)  
                                                                              \qquad \qquad   \qquad  \qquad \qquad \qquad \qquad \qquad \quad \qquad
                                \, \, E>\frac{k^2}{4}\qquad 
                                {\rm{odd \, solution}}  \nonumber
\end{eqnarray}
The quantities $\{ \psi_E, E \}$ are the eigenfunctions of the Schr\"odinger equation with potential:
\begin{eqnarray}
                            V_S(x)=\frac{h(x)^2}{4}+\frac{1}{2}\, \frac {\partial h(x)}{\partial x}= \frac{k^2}{4} - 4 \, \delta(x)  \label{RISKpot}
\end{eqnarray}
Hereafter, we will refer to this process as the $\delta$-process. The autocorrelation function of such process can be obtained in a closed form. In fact, by performing all integrations in Eq. \ref{ACRISKana} one gets:
\begin{eqnarray}
          &&   \hspace{-.5 cm} R^{(\delta)}(\tau)={2 \over k^2} (1 - 2 {\cal{T}}+4 {\cal{T}}^2+ {8 \over 3} {\cal{T}}^3)\,\Bigl(1-{\rm{Erf}}(\sqrt{{\cal{T}}})\Bigr)+  \label{ACrisken} \\
          &&   \quad -\,{{4 \sqrt{{\cal{T}}}}\over{3 k^2 \sqrt{\pi}}}\,(2 {\cal{T}}-1)(3 + 2 {\cal{T}})\,{\rm{exp}}(- {\cal{T}})                             \qquad 
               {\cal{T}}={k^2 \over 4} \tau \nonumber 
\end{eqnarray}
The result of Eq. \ref{ACRISKana} implies that the $\delta$-process is characterized by the presence of an infinite set of continuum timescales given by the inverse of the eigenvalues, Therefore such timescales are bounded from above, given that the possible eigenvalues are $E>k^2/4$. The autocorrelation function for large time lags behaves like a power-law with an exponential truncation associated to the lowest eigenvalue: $R(\tau) \approx {\rm{exp}}(-{k^2 \over 4} \tau) \tau^{-3/2}$ as $\tau \to \infty$.

The 2-point probability density function is given by \cite{risken,gardiner}
\begin{eqnarray}
                          && P^{(\delta)}(x_2, \tau; x_1, 0)=\psi_0(x_2)^2 \, \psi_0(x_1)^2 + \int_{k^2/4}^{\infty} \, dE \, \psi_0(x_2) \, \psi_E(x_2) \, \psi_0(x_1) \, \psi_E(x_1) \, e^{- E \tau} \label{congprobthRISK}
\end{eqnarray}
All integrations can be performed analytically, giving:
\begin{eqnarray}
                          && P^{(\delta)}(x_2, \tau; x_1, 0)=W_1(x_2,x_1,\tau) \theta(x_2) + W_2(x_2,x_1,\tau) \theta(-x_2) \label{congprobthRISKana} \\
                          && W_1(x_2,x_1,\tau)=W_1^+(x_2,x_1,\tau) \theta(x_1) + W_1^-(x_2,x_1,\tau) \theta(-x_1) \qquad
                                 W_2(x_2,x_1,\tau)=W_1(x_2,x_1,\tau) \nonumber \\
                          &&  W_1^+(x_2,x_1,\tau) =  {\rm{Erf}}\Bigl( \frac{x_2+x_1}{2\,\sqrt{\tau}}\Bigr) \,
                                                                            \frac{k^3 \, (x_1 +x_2-4/k)}{32} e^{-k(x_2+x_1)/2-k^2 \tau/4} + \nonumber \\
                          &&  \hspace{2.5 truecm}  {\rm{Erf}}\Bigl( \frac{x_2-x_1}{2\,\sqrt{\tau}}\Bigr) \,
                                                                            \frac{k^3 \, (x_1 -x_2)}{32} e^{-k(x_2+x_1)/2-k^2 \tau/4} +\nonumber \\
                          &&   \hspace{2.5 truecm}  \frac{k^2}{4} \, e^{-k (x_2+x_1)}  -
                                                                    \frac{k^3 \sqrt{\tau}}{16 \sqrt{\pi}} e^{-((x_2-x_1)^2+k^2 \tau^2+ 2 k \tau (x_2+x_1))/(4 \tau)}  + \nonumber \\
                          &&   \hspace{2.5 truecm}  \frac{k}{4 \sqrt{\pi} \sqrt{\tau}} e^{-(x_2-x_1+k \tau)^2/(4 \tau)}+
                                                                    \frac{k^3 \sqrt{\tau}}{16 \sqrt{\pi}} e^{-(x_2+x_1+k \tau)^2/(4 \tau)}        \nonumber \\                              
                     	 &&  W_1^-(x_2,x_1,\tau)  ={\rm{Erf}}\Bigl( \frac{x_2+x_1}{2\,\sqrt{\tau}}\Bigr) \,
                                                                            \frac{k^3 \, (x_2-x_1-4/k)}{32} e^{-k(x_2-x_1)/2-k^2 \tau/4} - \nonumber \\
                          &&   \hspace{2.5 truecm} {\rm{Erf}}\Bigl( \frac{x_2-x_1}{2\,\sqrt{\tau}}\Bigr) \,
                                                                            \frac{k^3 \, (x_1 +x_2)}{32} e^{-k(x_2-x_1)/2-k^2 \tau/4} + \nonumber \\
                          &&   \hspace{2.5 truecm}  \frac{k^2}{4} \, e^{k (x_2-x_1)}  -
                                                                    \frac{k^3 \sqrt{\tau}}{16 \sqrt{\pi}} e^{-((x_2+x_1)^2+k^2 \tau^2+ 2 k \tau (x_2-x_1))/(4 \tau)}  + \nonumber \\
                          &&   \hspace{2.5 truecm}  \frac{k}{4 \sqrt{\pi} \sqrt{\tau}} e^{-(x_2-x_1+k \tau)^2/(4 \tau)}+
                                                                    \frac{k^3 \sqrt{\tau}}{16 \sqrt{\pi}} e^{-(x_2-x_1+k \tau)^2/(4 \tau)}        \nonumber                
\end{eqnarray}

Again, we numerically integrated Eq. \ref{D1RISK} by considering a simple Euler scheme with $x_0=0.1$ and $k=0.5$. In Table \ref{tab:ACRISK} we show the autocorrelation function values of such process obtained at different lags (first column) by using the analytical expression of Eq. \ref{ACrisken} (second column) and the new computing method of Eq. \ref{ACnew} (third column), with $n=100$ and $\tau_M=50$. The values shown in the table refer to a single realization of the process with $x_m=-23.764$, $x_M=24.377$ and therefore $\Delta=0.481$. The table again confirms the correctness of Eq. \ref{ACnew}.
\begin{table} [H]
\small
\caption{$\delta$ process. Autocorrelation function values of the $\delta$ process obtained at different lags (first column) by using the analytical expression of Eq. \ref{ACrisken} (second column) and the new computing method of Eq. \ref{ACnew} (third column), with $n=100$ and $\tau_M=50$. The values shown in the table refer to a single realization of the process with $x_m=-23.764$, $x_M=24.377$ and therefore $\Delta=0.481$.}
\begin{tabular}{ccc}
\textbf{lag}	& \textbf{analytical}	& \textbf{new}	\\
 1 & 7.09226 & 7.16095 \\
 2 & 6.32512 & 6.39185 \\
 3 & 5.66376 & 5.72534 \\
 4 & 5.08733 & 5.14499 \\
 5 & 4.58115 & 4.63344 \\
 6 & 4.13415 & 4.18142 \\
 7 & 3.73765 & 3.78116 \\
 8 & 3.38465 & 3.42476 \\
 9 & 3.06939 & 3.10428 \\
 10 & 2.78709 & 2.82 \\
 11 & 2.53372 & 2.56558 \\
 12 & 2.30585 & 2.33737 \\
 13 & 2.10053 & 2.1305 \\
 14 & 1.91522 & 1.94051 \\
 15 & 1.74773 & 1.77059 \\
 16 & 1.59613 & 1.61735 \\
 17 & 1.45875 & 1.47558 \\
 18 & 1.3341 & 1.34741 \\
 19 & 1.22089 & 1.23164 \\
 20 & 1.11797 & 1.12635 \\
 21 & 1.02431 & 1.03268 \\
 22 & 0.939005 & 0.94803 \\
 23 & 0.861248 & 0.866215 \\
 24 & 0.790317 & 0.792533 \\
 25 & 0.725566 & 0.724728 \\
 26 & 0.666416 & 0.663397 \\
 27 & 0.612348 & 0.608951 \\
 28 & 0.562895 & 0.558124 \\
 29 & 0.517638 & 0.511327 \\
 30 & 0.476197 & 0.469649 \\
 31 & 0.43823 & 0.433253 \\
 32 & 0.403429 & 0.399852 \\
 33 & 0.371515 & 0.370159 \\
 34 & 0.342235 & 0.34282 \\
 35 & 0.315359 & 0.317417 \\
 36 & 0.290679 & 0.294129 \\
 37 & 0.268008 & 0.274497 \\
 38 & 0.247173 & 0.255192 \\
 39 & 0.228019 & 0.236703 \\
 40 & 0.210404 & 0.222586 \\
 41 & 0.194198 & 0.210434 \\
 42 & 0.179284 & 0.198864 \\
 43 & 0.165554 & 0.190179 \\
 44 & 0.15291 & 0.180624 \\
 45 & 0.141264 & 0.170356 \\
 46 & 0.130532 & 0.158773 \\
 47 & 0.120641 & 0.147435 \\
 48 & 0.111522 & 0.138426 \\
 49 & 0.103113 & 0.128428 \\
 50 & 0.0953561 & 0.117806 \\
\end{tabular}
\label{tab:ACRISK}
\end{table}

\begin{figure}[H]
\begin{center}
{\includegraphics[scale=0.65]{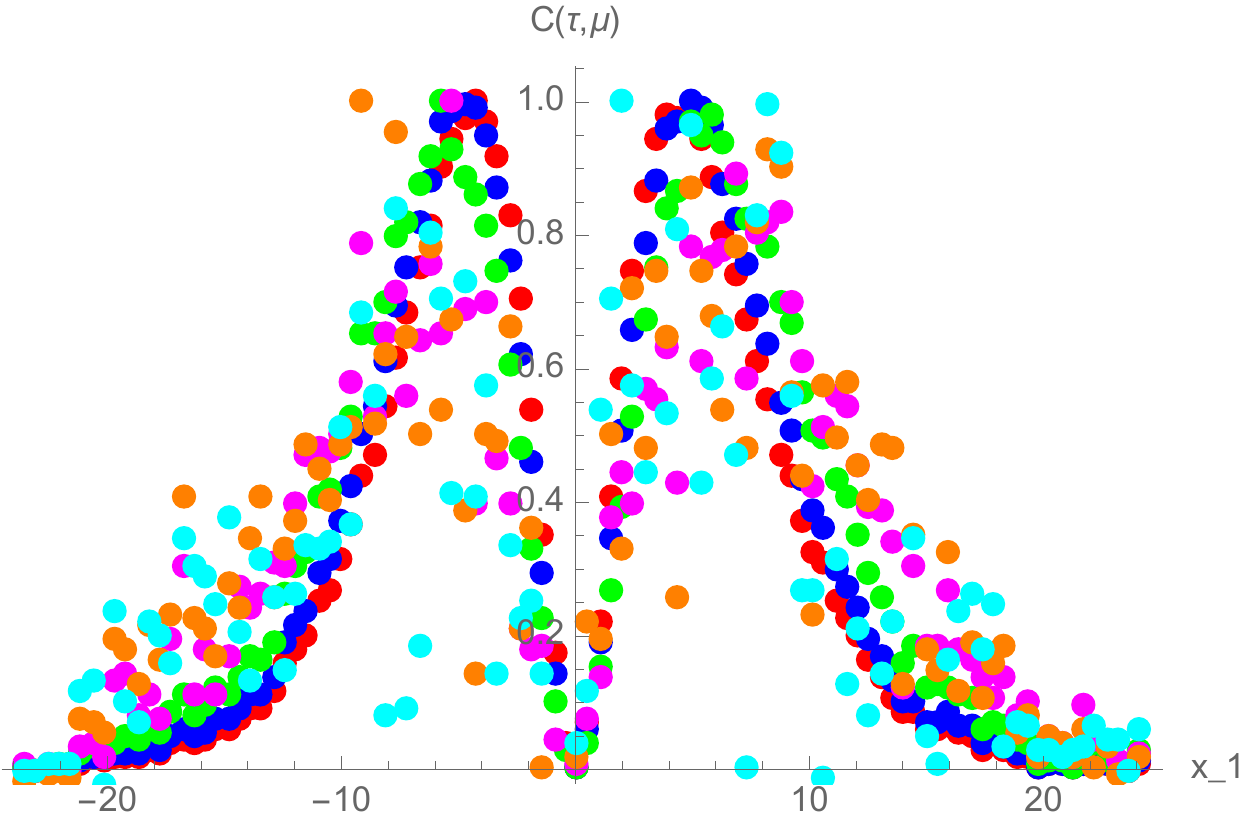}
 \includegraphics[scale=0.65]{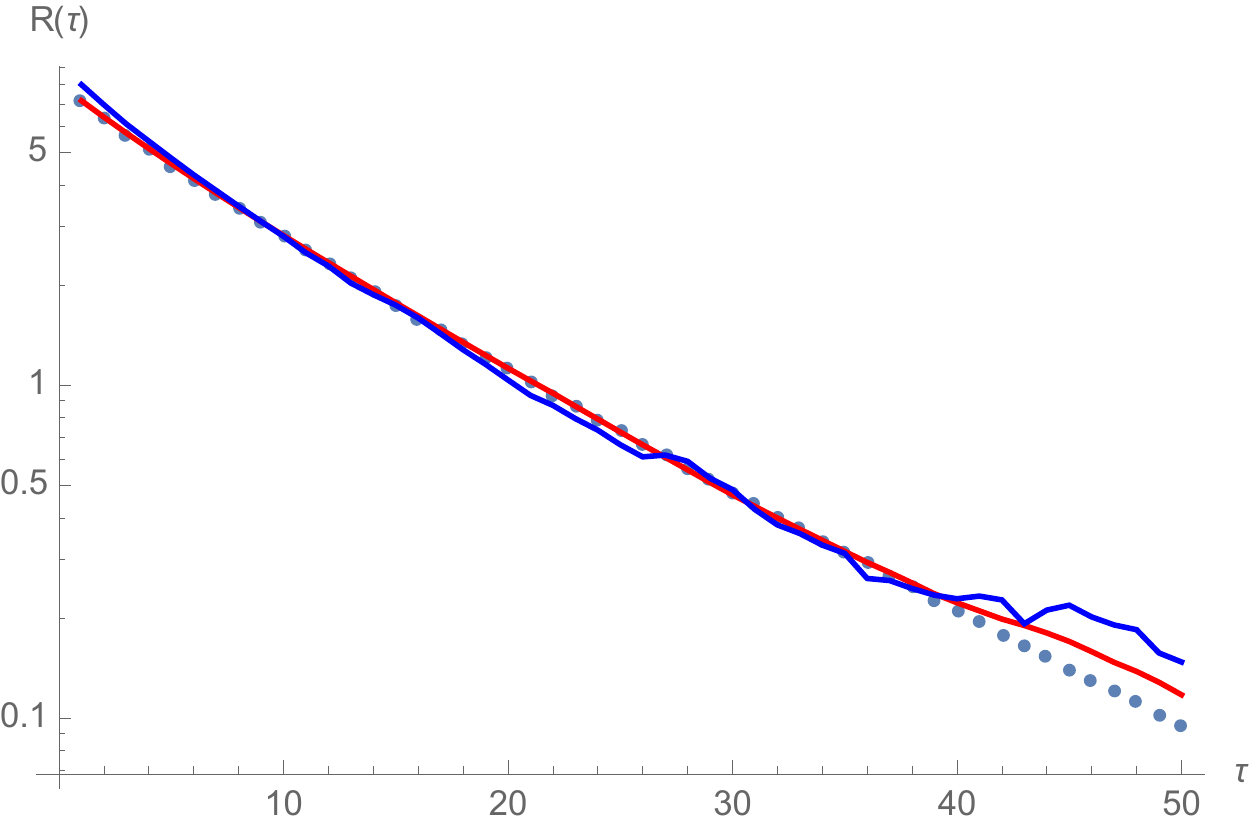}}
\end{center}
\caption{$\delta$ process. The left panel of Fig. \ref{contrib_RISK} shows the quantities $C^{(\delta)}(\tau,\mu)$ rescaled with their maximal value for different value of $\tau$: $\tau=1$ (red), $\tau=3$ (blue), $\tau=10$ (green), $\tau=30$ (magenta), $\tau=40$ (orange), $\tau=50$ (cyan). The blue line in the right panel of Fig.  \ref{contrib_RISK} shows the maximal values of the $C^{(\delta)}(\tau,\mu)$ curves shown in the left panel, i.e. the factors that have been used to normalize that curves. The red line in this panel corresponds to the values in the third column of Table \ref{tab:ACRISK}, while the circles correspond to the theoretical prediction given by second column of Table \ref{tab:ACRISK}. The right panel is in a log-linear scale.}\label{contrib_RISK}
\end{figure}

In the left panel of Fig. \ref{contrib_RISK} we show the quantities $C^{(\delta)}(\tau,\mu)$ rescaled with their maximal value for different value of $\tau$: $\tau=1$ (red), $\tau=3$ (blue), $\tau=10$ (green), $\tau=30$ (magenta), $\tau=40$ (orange), $\tau=50$ (cyan). The blue line in the right panel of Fig.  \ref{contrib_RISK} shows the maximal values of the $C^{(\delta)}(\tau,\mu)$ curves shown in the left panel, i.e. the factors that have been used to normalize those curves. The red line in this panel corresponds to the values in the third column of Table \ref{tab:ACRISK}, while the circles correspond to the theoretical prediction given by second column of Table \ref{tab:ACRISK}. The six curves in the left panel no longer collapse onto a single curve. This is essentially due to the fact that this process is genuinely multiscale, i.e. the autocorrelation function is not approximable with an exponential function. However, the right panel shows that the maxima of the contributions $C^{(\delta)}(\tau,\mu)$ are still a good proxy of $R^{(\delta)}(\tau)$. Let us now consider the theoretical counterparts of the contributions: 
\begin{eqnarray}
                           C_{th}^{(\delta)}=\Delta^2 \, T \, \int_{-\infty}^{+\infty} \, dx_1 \, x \, x_1 \, P^{(\delta)}(x, \tau; x_1, 0) \label{contribthRISK}
\end{eqnarray}
All integrations can be performed analytically, giving:
\begin{eqnarray}
                          && C_{th}^{(\delta)}(x, \tau)=\Delta^2 \, T \, \Bigl( C^+(x,\tau) \theta(x) + C^-(x,\tau) \theta(-x) \Bigr) \label{ccontribthRISKana} \\
                          &&  C^+(x,\tau)  = \frac{k \, x}{4} \, e^{- k x} \, 
                                                           \bigl( x - k \tau
                                                           \bigl)  \,
                                                           {\rm{Erfc}}(\frac{k}{2} \sqrt{\tau} - \frac{x}{2 \, \sqrt{\tau}}) +
                                                            \frac{k \, x}{4} \, 
                                                           \bigl( x + k \tau
                                                           \bigl)  \,
                                                           {\rm{Erfc}}(\frac{k}{2} \sqrt{\tau} + \frac{x}{2 \, \sqrt{\tau}}) \nonumber      \\
                          &&  C^-(x,\tau) = \frac{k \, x}{4} \,  
                                                           \bigl( x - k \tau
                                                           \bigl)  \,
                                                           {\rm{Erfc}}(\frac{k}{2} \sqrt{\tau} - \frac{x}{2 \, \sqrt{\tau}}) +
                                                            \frac{k \, x}{4} \, e^{+ k x} \,
                                                           \bigl( x + k \tau
                                                           \bigl)  \,
                                                           {\rm{Erfc}}(\frac{k}{2} \sqrt{\tau} + \frac{x}{2 \, \sqrt{\tau}}).    \nonumber
\end{eqnarray}
By using the above expression it is possible to show that the maxima of $C^{(\delta)}(x,\tau)$ depend on $\tau$. We can not therefore factorize $C^{(\delta)}(x, \tau)$ into a spatial and a temporal term. However, at fixed time lag, the contributions show a behaviour qualitatively similar to the one observed for the OU process, thus indicating that also in this case, the main contributions to the autocorrelation function come from large $x(t)$ values.

Let us now introduce $C_{th,L}^{(\delta)}=C_{th}^{(\delta)} \, H(L-x) \, H(L+x)$, where $H(\cdot)$ is the Heaviside function. By performing the relevant integrations we get:
\begin{eqnarray}
                    &&       R_L^{(\delta)}(\tau)=q_0+q_1 \, e^{- k L} \Biggl( 1+ {\rm{Erf}}\biggl( \frac{L - k \tau}{2 \sqrt{\tau}}\biggl) \Biggl) + 
                                                                q_2 \, {\rm{Erf}}\biggl( \frac{L + k \tau}{2 \sqrt{\tau}}\biggl) + 
                                                                q_3 \, e^{-\frac{(L + k \tau)^2}{4 \tau}}+ 
                                                                q_4 \, {\rm{Erf}}\biggl( \frac{k \sqrt{\tau}}{2 }\biggl)+ 
                                                                q_5 \, e^{-\frac{k^2 \tau}{4}} \\
                    &&       q_0= \frac{1}{k^2} -\frac{\tau}{2} + \frac{k L^3}{6} +\frac{k^2 L^2 \tau}{4} \qquad \qquad  \qquad \qquad \qquad 
                                q_1= -\frac{1}{k^2} -\frac{L}{k} +\frac{\tau - L^2 + k L \tau}{2} \nonumber \\
                    &&       q_2= \frac{1}{k^2} -\frac{k L^3}{6} -\frac{\tau}{2} -\frac{k^2 L^2 \tau}{4} -\frac{k^2 \tau^2}{2} -\frac{k^4 \tau^3}{12} \qquad \quad
                                q_3=\frac{\sqrt{\tau}}{6 k \sqrt{\pi}} \Bigl( -12 -  6 k L - k^3 L \tau  + k^4 \tau^2  - 2 k^2 (L^2 - 2 \tau) \Bigr)\nonumber \\
                    &&       q_4=-\frac{2}{k^2} + \tau - \frac{k^2 \tau^2}{2} - \frac{k^4 \tau^3}{12} \qquad \qquad \qquad \qquad \qquad 
                                q_5=\frac{2 \sqrt{\tau}}{k \sqrt{\pi}} - \frac{2 k \tau^{3/2}}{3 \sqrt{\pi}} - \frac{k^3 \tau^{5/2}}{6 \sqrt{\pi}}\nonumber
 \end{eqnarray}
This expression shows how the $R_L^{(\delta)}(\tau)$ converges to $R^{(\delta)}(\tau)$  due to the $e^{- k L}$ term. Other terms of the order $e^{-L^2/(4 \tau)}$ are present, but they decay faster. Indeed, Eq. \ref{ccontribthRISKana} indicates that for large $x_1$ values the behaviour of $C_{th}^{(\delta)}(x_1, \tau)$ is dominated by the exponential term $e^{- k x_1}$ which is ultimately related to the tail of the stationary pdf $|\psi_0(x)|^2$. 

%%%%%%%%%%%%%%%%%%%%%%%%%%%%%%%%%%%%%%%%%%%%%%%%%%%%%%%%%%%%%%%%%%%%%%%%%%%%%%%%%%%%%%%%%%%%%%%%%%%%
%%%%%%%%%%%%%%%%%%%%%%%%%%%%%%%%%%%%%%%%%%%%%%%%%%%%%%%%%%%%%%%%%%%%%%%%%%%%%%%%%%%%%%%%%%%%%%%%%%%%
%\section{Power-law correlated process} \label{ARFIMA}
%%%%%%%%%%%%%%%%%%%%%%%%%%%%%%%%%%%%%%%%%%%%%%%%%%%%%%%%%%%%%%%%%%%%%%%%%%%%%%%%%%%%%%%%%%%%%%%%%%%%
%%%%%%%%%%%%%%%%%%%%%%%%%%%%%%%%%%%%%%%%%%%%%%%%%%%%%%%%%%%%%%%%%%%%%%%%%%%%%%%%%%%%%%%%%%%%%%%%%%%%

%%%%%%%%%%%%%%%%%%%%%%%%%%%%%%%%%%%%%%%%%%%%%%%%%%%%%%%%%%%%%%%%%%%%%%%%%%%%%%%%%%%%%%%%%%%%%%%%%%%%
%\subsection{A gaussian power-law correlated process} \label{chimeragauss}
%%%%%%%%%%%%%%%%%%%%%%%%%%%%%%%%%%%%%%%%%%%%%%%%%%%%%%%%%%%%%%%%%%%%%%%%%%%%%%%%%%%%%%%%%%%%%%%%%%%%

%%%%%%%%%%%%%%%%%%%%%%%%%%%%%%%%%%%%%%%%%%%%%%%%%%%%%%%%%%%%%%%%%%%%%%%%%%%%%%%%%%%%%%%%%%%%%%%%%%%%
%\subsection{A power-law distributed process with power-law decaying integrable autocorrelation} \label{chimeraint}
%%%%%%%%%%%%%%%%%%%%%%%%%%%%%%%%%%%%%%%%%%%%%%%%%%%%%%%%%%%%%%%%%%%%%%%%%%%%%%%%%%%%%%%%%%%%%%%%%%%%

%%%%%%%%%%%%%%%%%%%%%%%%%%%%%%%%%%%%%%%%%%%%%%%%%%%%%%%%%%%%%%%%%%%%%%%%%%%%%%%%%%%%%%%%%%%%%%%%%%%%
%\subsection{A power-law distributed process with power-law decaying not-integrable autocorrelation} \label{chimeranoint}
%%%%%%%%%%%%%%%%%%%%%%%%%%%%%%%%%%%%%%%%%%%%%%%%%%%%%%%%%%%%%%%%%%%%%%%%%%%%%%%%%%%%%%%%%%%%%%%%%%%%

%%%%%%%%%%%%%%%%%%%%%%%%%%%%%%%%%%%%%%%%%%%%%%%%%%%%%%%%%%%%%%%%%%%%%%%%%%%%%%%%%%%%%%%%%%%%%%%%%%%%
%%%%%%%%%%%%%%%%%%%%%%%%%%%%%%%%%%%%%%%%%%%%%%%%%%%%%%%%%%%%%%%%%%%%%%%%%%%%%%%%%%%%%%%%%%%%%%%%%%%%
\section{Discussion and Conclusions} \label{concl}
%%%%%%%%%%%%%%%%%%%%%%%%%%%%%%%%%%%%%%%%%%%%%%%%%%%%%%%%%%%%%%%%%%%%%%%%%%%%%%%%%%%%%%%%%%%%%%%%%%%%
%%%%%%%%%%%%%%%%%%%%%%%%%%%%%%%%%%%%%%%%%%%%%%%%%%%%%%%%%%%%%%%%%%%%%%%%%%%%%%%%%%%%%%%%%%%%%%%%%%%%

By considering explicit examples in sections \ref{OU}, \ref{SW} and \ref{RISK}, we have shown that Eq. \ref{ACnew} indeed provides an alternative way of computing the autocorrelation function of stationary stochastic processes.

The advantage of using the approach of Eq. \ref{ACnew} is twofold. On one side, we have seen that the knowledge of ${\cal{N}}(\tau,g_\mu,g_\nu)$ immediately gives information about the auto-covariance function $R(\tau)$ and the histogram $H(\mu)$ of the considered process. On the other hand, starting from ${\cal{N}}(\tau,g_\mu,g_\nu)$, one can compute $C(\tau,\mu)$, as defined in Eq. \ref{ACcontribdef}. This quantity allows to quantitatively understand which parts of the numerical series most contribute to the autocorrelation function.  In fact, we have investigated what is the contribution of large $x$ values to the autocorrelation function by considering the truncated functions $C_{L}(\tau,\mu)=C(\tau,\mu) \, H(L-x) \, H(L+x)$. Expectedly, the tails of  $C(\tau,\mu)$ tell us how crucial is the contribution of large process values to the numerical evaluation of the autocorrelation function. However, we have also shown that for the specific processes considered here the tails of $C(\tau,\mu)$ are intimately related to the tails of the stationary pdf $|\psi_0|^2$ and therefore for such processes we can conclude that the pdf tails tell us how crucial is the contribution of large process values to the numerical evaluation of the autocorrelation function. Indeed, this result can be generalized to all stochastic processes admitting a nonlinear Langevin equation with additive noise for which Eq. \ref{congprobthSW} and Eq. \ref{congprobthRISK}  hold true \cite{myuno}. In fact, one can write: 
\begin{eqnarray}
                     &&      C_{th} = x \psi_0(x) \int dE \, e^{- E t} \, c_E \, \psi_E(x)\nonumber \\
                     &&      c_E= \int_{-\infty}^{+\infty} \, dx_1 \, x_1 \, \psi_0(x_1) \, \psi_E(x_1)
\end{eqnarray}
For large values $\psi_E(x) \approx A_E e^{- i \sqrt{E} x}$ for any stochastic process admitting a quantum potential $V_S(x)$ well behaved at infinity.  Therefore, for large $x$ values we get $C_{th} \propto x \psi_0(x)$, which is in agreement with all the above results. 

When considering stochastic processes with power-law pdf tails $\psi_0(x) \propto x^{-\alpha/2}$, one gets $C_{L}(\tau,\mu) \propto 1/L^{\alpha/2-1}$, thus indicating that the convergence to the theoretical prediction can be very slow or, conversely, that the {\em{error}} can be not negligible. When considering stochastic processes $\xi(t)$ that can be obtained as a coordinate transformation $\xi=f(x)$ starting from a   processes $x(t)$ admitting a nonlinear Langevin equation with additive noise for which Eq. \ref{congprobthSW} and Eq. \ref{congprobthRISK}  hold true, one can prove that $C_{th} \approx f(x) \psi_0(x)$ \cite{mydue}. Also in this case the role of the pdf tail is crucial.  While the protocol of Eq. \ref{ACnew} is valid for any stochastic process, further work is needed to confirm whether such role of the pdf also holds true for processes not described by Langevin equations.

The study of  ${\cal{N}}(\tau,g_\mu,g_\nu)$ also helps in characterizing processes with multiple time-scales. In all considered cases, for each time lag $\tau$, the quantity  $C(\tau,\mu)$ has a peculiar bimodal shape. Our results in sections \ref{OU} and \ref{SW} show that when the tail of $R(\tau)$ is well described by an effective exponential function, i.e. the process exhibits a single timescale, then the location of these peaks is independent of the time lag $\tau$. Therefore  $C(\tau,\mu)$ can be approximated by its factorized form $C(\tau,\mu) \approx F(x) {\cal{R}}(\tau)$ where numerical evidences show that  ${\cal{R}}(\tau)\approx R(\tau)$. In the case when the stochastic process is a truly multiscale one, i.e. its autocorrelation function can not be approximated by an exponential function, then the above factorization is no longer observed, as in section \ref{RISK}. 
%However, for each time lag $\tau$, $C(\tau,\mu)$ still has a peculiar bimodal shape suggesting that the bins associated to low of very high $x$ values poorly contribute to the autocorrelation function. 
%Section \ref{ARFIMA} shows that for power-law  correlated or even long-range correlated processes $C(\tau,\mu)$ exhibits the same behaviour as the one observed in section \ref{RISK}. XXXXXXXXXXXXXXXX

From a theoretical point of view, the importance of the present approach consists in setting a bridge between the time-average computation and the ensemble average computation of the autocovariance function. In fact, ensemble average simulations of the autocovariance function require the evaluation of the quantity:
\begin{eqnarray}
                          &&  R_{EA}(\tau)=\langle X(\tau) \, X(0) \rangle = \frac{1}{N} \, \sum_{r=1}^{N} \, x_0^{(r)} \, x_{\tau}^{(r)} = \int_{-\infty}^{+\infty} dx_1 \int_{-\infty}^{+\infty} dx_2 \, x_1 \, x_2 \, P(x_2,\tau; x_1,0) \label{ACens}
\end{eqnarray}
where $x_t^{(r)}$ is the value of the process at the $i$-th event of the $r-$th realization while $N$ is the number of process realizations. When the process is ergodic $R_{TA}(\tau)$ of Eq. \ref{ACtime} and $R_{EA}(\tau)$ of Eq. \ref{ACens} give the same result and therefore one gets:
\begin{eqnarray}
                          &&  \frac{1}{T} \sum_{\mu=1}^n  \Bigl( \frac{g_{\mu-1}+g_\mu}{2}\Bigl)   \, 
                                                              \sum_{\nu=1}^n  \Bigl( \frac{g_{\nu-1}+g_\nu}{2}\Bigl) \, 
                                                              {\cal{N}}(\tau,g_\mu,g_\nu) =
                               %\overline{X(0) \, X(\tau)}=\langle X(\tau) \, X(0) \rangle = 
                               \int_{-\infty}^{+\infty} dx_1 \int_{-\infty}^{+\infty} dx_2 \, x_1 \, x_2 \, P(x_2,\tau; x_1,0) \label{ACergo}
\end{eqnarray}
With the approach of Eq. \ref{ACnew} we essentially evaluate the 2-point joint probability  $P(x_2,\tau; x_1,0)$ starting form a single realization of the process and by counting the number ${\cal{N}}(\tau,g_\mu,g_\nu)$ of positive outcomes. 

Finally, another important issue regards the possibility of investigating the memory properties of the process by using the conditional entropy approach \cite{condentro}:
\begin{eqnarray}
                          && S(x_2,\tau| x_1,0)= - \int d x_1 \int d x_2 \,  P(x_2,\tau; x_1,0) \log \Bigl( 
                                                                                                                                              \frac{P(x_2,\tau; x_1,0)}{P(x_1,0)}
                                                                                                                                          \Bigr)   \label{CondEntro}
\end{eqnarray}
i.e. by computing the expectation value of the log conditional probability. 

The above results call for future work devoted to a deeper investigation of the interconnections between an autocorrelation-based and a conditional-entropy-based study of the memory processes of stochastic processes, also in relation with their ergodic properties.

\bibliographystyle{IEEEtran}

\end{document}